\documentclass[a4paper,11pt]{article}
\usepackage{jheppub}
\usepackage[FIGTOPCAP]{subfigure}
\usepackage{amsmath}
\usepackage{amsfonts}
\usepackage{amssymb}
\usepackage{tensor,epigraph}
\usepackage{cancel}
\usepackage{ulem} 
\usepackage{musixtex}
\usepackage{booktabs}
\usepackage{empheq}
\usepackage{amssymb}
\usepackage{amsthm}
\usepackage{comment}
\usepackage{psfrag}
\usepackage{graphicx}
\usepackage{color}
\usepackage{bbm}
\usepackage{color}
\usepackage{natbib}
\usepackage[table]{xcolor}

\newcommand{\be}{\begin{equation}}
\newcommand{\ee}{\end{equation}}

\definecolor{klgreen}{rgb}{0.0, 0.5, 0.0}

\newcommand{\exclude}[1]{}

\newcommand{\beq}{\begin{equation}}
\newcommand{\eeq}{\end{equation}}
\newcommand{\bea}{\begin{eqnarray}}
\newcommand{\eea}{\end{eqnarray}}
\long\def\/*#1*/{}

\newcommand{\junk}[1]{}

\title{ $\mathbb{T}^{2}$- inflation: Sourced by energy-momentum squared gravity}
\author[a]{Seyed Ali Hosseini Mansoori}
\author[a]{, Fereshteh Felegary}
\author[b,c]{, Mahmood Roshan}
\author[d]{, \"{O}zg\"{u}r Akarsu}
\author[e,f,g]{, and Mohammad Sami }

\affiliation[a]{Faculty of Physics, Shahrood University of Technology,\\
P.O. Box 3619995161, Shahrood, Iran}
\affiliation[b]{Department of Physics, Faculty of Science, Ferdowsi University of Mashhad,\\
P.O. Box 1436, Mashhad, Iran}
\affiliation[c]{School of Astronomy, Institute for Research in Fundamental Sciences (IPM), Tehran, Iran, P.O. Box 19395-5531}
\affiliation[d]{Department of Physics, Istanbul Technical University,\\
Maslak 34469 Istanbul, Turkey}
\affiliation[e]{Centre for Cosmology and Science Popularization (CCSP), SGT University,\\
Gurugram, Delhi- NCR, Haryana- 122505, India}
\affiliation[f]{Eurasian International Centre for Theoretical Physics, Astana, Kazakhstan}
\affiliation[g]{Chinese Academy of Sciences,\\
52 Sanlihe Rd, Xicheng District, Beijing}

\emailAdd{shosseini@shahroodut.ac.ir}
\emailAdd{fereshteh.felegary@gmail.com}
\emailAdd{mroshan@um.ac.ir}
\emailAdd{akarsuo@itu.edu.tr}
\emailAdd{ sami\_ccsp@sgtuniversity.org}

\abstract{In this paper, we examine chaotic inflation within the context of the energy-momentum squared gravity (EMSG) focusing on the energy-momentum powered gravity (EMPG) that incorporates the functional $f(\mathbb{T}^2)\propto (\mathbb{T}^2)^{\beta}$ in the Einstein-Hilbert action, in which $\beta$ is a constant and $\mathbb{T}^2\equiv T_{\mu \nu}T^{\mu \nu}$ where $T_{\mu \nu}$ is the energy-momentum tensor, which we consider to represent a single scalar field with a power-law potential. We also demonstrate that the presence of EMSG terms allows the single-field monomial chaotic inflationary models to fall within current observational constraints, which are otherwise disfavored by Planck and BICEP/Keck findings. We show that the use of a non-canonical Lagrangian with chaotic potential in EMSG can lead to significantly larger values of the non-Gaussianity parameter, $f_{\rm Nl}^{\rm equi}$ whereas EMSG framework with canonical Lagrangian gives rise to results similar to those of the standard single-field model.

}
\begin{document}

\maketitle
\section{Introduction}
Over the past few years, a variety of extended theories of gravity have been discussed in the literature, motivated by theoretical and observational considerations. For instance, some of these schemes aim to explain the cosmic speed-up, whereas others propose to replace the role of dark matter with a modification of gravity. There are several theories that have been proposed to extend General Relativity (GR) by adding new gravitational scalar fields to the Einstein-Hilbert action. Some of these theories include the Brans-Dicke scalar-tensor theory~\cite{BD} and the mimetic dark matter theory~\cite{mimetic}. To address the dark matter problem, some theories also incorporate an additional vector field; examples of such theories include TeVeS~\cite{teves}, MOG~\cite{mog} and the new relativistic theory of Modified Newtonian Dynamics (MOND)~\cite{skordis}. While some theories introduce new fields to extend General Relativity (GR), others modify the existing fields instead. An example of the latter is $f(R)$ gravity where a function of the Ricci scalar $R$ is used in the action~\cite{faraoni}. Although this theory can be transformed into a scalar-tensor theory, the theory itself only involves the metric tensor as the gravitational field. In non-local gravity (NLG)~\cite{mashhoon}, the metric tensor is similarly the only gravitational field present. 

In addition, there exists a specific class of modified theories which permit the presence of scalars constructed from the energy-momentum tensor $T_{\mu\nu}$ in the action.  An illustration of this concept can be seen in $f(R,T)$ gravity, where the action involves the scalar $T=g^{\mu\nu}T_{\mu\nu}$, which is the trace of $T_{\mu\nu}$~\cite{harko}. Similarly, the $f(R,\mathbb{T}^2)$ model includes an arbitrary function, $\mathbb{T}^2\equiv T^{\mu\nu}T_{\mu\nu}$ (viz., the self-contraction of the energy-momentum tensor $T^{\mu\nu}$) in the action~\cite{Katirci,roshan2016energy,akarsu2017,board2017}. This model is commonly known in the literature as Energy-Momentum-Squared-Gravity (EMSG). 

Before we proceed further, let us briefly review the current status of EMSG within the wider landscape of modified gravity theories. Numerous studies in the literature have delved into the cosmological and astrophysical consequences of EMSG. For instance, in~\cite{roshan2016energy}, a basic model of the form $f(R,\mathbb{T}^2)=R-\alpha \mathbb{T}^2$ is employed to investigate bouncing cosmological solutions and address the big bang singularity. Additionally, the EMSG incorporates additional terms that introduce quadratic pressure and density terms into the Friedmann equations, similar to the corrections found in loop-quantum gravity~\cite{ashtekar} for $\alpha>0$, and the brane world scenarios~\cite{Brax:2003fv} for $\alpha<0$. These terms, for $\alpha>0$, permit the possibility of bouncing solutions. After conducting further investigations, doubts were raised about the feasibility of the cosmological bounce in the aforementioned simple model~\cite{barbar}. However, in Ref.~\cite{sarvi}, the authors successfully obtained feasible bouncing solutions in EMSG by introducing the Palatini version of the theory. 

In the framework of EMSG, it is possible to achieve late-time accelerated expansion through conventional matter-energy sources without having to incorporate the cosmological constant; for instance, in the EMSG of the form $f(R,\mathbb{T}^2)= R-\alpha (\mathbb{T}^2)^{\beta}$ [known also as energy-momentum powered gravity (EMPG)], provided that $\beta\sim0$~\cite{akarsu2017,board2017}. To explore various cosmological exact solutions in EMPG, we recommend referring to~\cite{board2017}. The dynamical systems analysis of EMSG has also been performed in~\cite{roshan2016energy,bahamonde}, which demonstrates that the simple $R-\alpha \mathbb{T}^2$ model features a suitable series of cosmological fixed points (or epochs). Furthermore,~\cite{akarsu2,akarsu4} conduct a comprehensive analysis of the cosmological implications of the scale-independent EMSG, specifically with regard to the $f(R,\mathbb{T}^2)= R-\alpha \sqrt{\mathbb{T}^2}$ term; which can also lead to the accelerated expansion of the universe. Last but not least, it should be noted that most of the modifications to gravity involve extra degrees of freedom, whereas EMSG does not resort to the same.

Several studies have also explored the astrophysical implications of EMSG, including the post-Newtonian limit of EMSG and the bending of light experiment analyzed in~\cite{nazari}, the structure of compact stars in EMSG detailed in~\cite{nari}, the Jeans analysis in EMSG conducted in~\cite{kazemi}, constraints on EMSG from binary pulsar observations explained in~\cite{nazari3,nazari2}, and constraints on EMSG from neutron star observations discussed in~\cite{akarsu3}. Additionally, certain cosmological observations have imposed limitations on the free parameter of EMSG, namely $\alpha$~\cite{faraji,Ranjit:2020syg}.

 On the other hand, in the framework of inflationary cosmology, the recent Planck results put severe restriction on the inflationary parameters. For instance, using observations from Planck, WMAP, and BICEP/Keck during the 2018 observing season~\cite{ade2021improved}, the tensor-to-scalar ratio parameter is limited to $r<0.036$ at $95\%$ confidence. As a result of this limitation, chaotic inflation~\cite{linde1983chaotic,linde1982new} with a potential of $\phi^{n}$, even for $n=2/3$, has been excluded at about $95\%$ CL. Therefore, the focus of our paper is to examine the inflationary parameters - such as the spectral index $n_s$, tensor-to-scalar ratio $r$, and non-Gaussianity \cite{Maldacena:2002vr,Maldacena:2011nz,Choudhury:2017glj,Choudhury:2012whm,Choudhury:2014uxa,Celoria:2018euj,Chen:2010xka,Baumann:2009ds,Senatore:2016aui,Baumann:2018muz} parameter in an equilateral shape $f_{\rm Nl}^{\rm equi}$ - for chaotic inflation in the presence of EMSG corrections. Our expectation is that the EMSG corrections will yield values of $n_{s}$ and $r$ that fall within the current BICEP/Keck bound~\cite{aghanim2020planck,ade2021improved}, as opposed to being ruled out in the standard model of chaotic inflation.

The rest of the paper is organized as follows: In Section \ref{sec2}, we attempt to construct our model within the EMGS framework by using the energy-momentum tensor associated with the canonical single field Lagrangian. In this regard, this setup is a subset of K-essence models~\cite{armendariz2001essentials}. Section \ref{sec3} begins by examining the stability of the model at the level of cosmological perturbations. In particular, this finding imposes restrictions on the free parameter of EMSG, disfavoring many works have been done in the EMSG model. In Section \ref{sec4}, under the slow-roll scheme, we then  obtain an inflationary solution via the background solutions in our scenario. Moreover, the inflationary parameters - such as $\{n_s, r, f_{\rm Nl}^{\rm equi}\}$, are both analyzed and discussed, especially at the end of Section \ref{sec4}. These findings are compared to those of standard chaotic inflation~\cite{linde1983chaotic,linde1982new}. One striking feature of EMSG corrections is that they shift the tensor-to-scalar ratio $r$ to smaller values, which brings them in line with the recent BICEP/Keck bound. Furthermore, in Section \ref{sec5}, we determine the inflationary parameters by considering the energy-momentum tensor derived from a non-canonical Lagrangian~\cite{Li:2012vta}. The value of $f_{\rm Nl}^{\rm equi}$ is significantly larger when compared to that in the canonical Lagrangian. Our conclusions are drawn in Section \ref{consec}.

\section{The model and  background field equations}\label{sec2}
Let us start by taking the EMPG model described by the following action~\cite{akarsu2017,board2017}:
\begin{equation}\label{action}
S = \frac{1}{2}\int d^{4}x \sqrt{-g}\left[ M_{\text{\rm p}}^2 R  - \alpha\, M_{\rm p}^{4(1-2 \beta)} (\mathbb{T}^{2})^{\beta}+2\mathcal{L}_{\rm m}\right],
\end{equation}
where $M_{\rm p}$ is the reduced Planck mass, $R$ is the Ricci scalar associated with the spacetime metric $g_{\mu \nu}$,  $\mathcal{L}_{\rm m}$ is the Lagrangian density corresponding to the matter source described by the energy-momentum tensor $T_{\mu \nu}$. In addition, $\mathbb{T}^2\equiv T_{\mu \nu}T^{\mu \nu}$ is a scalar and $\alpha$ is a dimensionless constant that determines the coupling strength of the EMPG modification. It should be stressed that in~\cite{roshan2016energy}, the specific case $\beta=1$ is explored. So here we study a more general case.

Unlike previous studies, such as~\cite{harko,Katirci,roshan2016energy,Board:2017ign}, where the perfect fluid energy-momentum tensor $T_{\mu\nu}= (\rho+p) u_{\mu} u_{\nu}+pg_{\mu \nu}$ (where $\rho$ is the energy density, $p$ is the thermodynamic pressure, and $u_{\mu}$ is the four-velocity satisfying the conditions $u_{\mu}u^{\mu}=-1$\footnote{We use the metric signature, (-,+,+,+).}) was used, in this work we construct $T_{\mu \nu}$ by varying the canonical scalar field Lagrangian $\mathcal{L}_{\rm m}=X-V(\phi)$ where $X=-(\partial_{\mu} \phi \partial^{\mu} \phi)/2$ with respect to the metric. In this case, $T_{\mu\nu}$ reads
\begin{equation}\label{SM}
T_{\mu \nu}\equiv -\frac{2}{\sqrt{-g}} \frac{\delta(\sqrt{-g} \mathcal{L}_{\rm m})}{\delta g^{\mu \nu}}=\partial_{\mu} \phi \partial_{\nu} \phi + g_{\mu \nu} ( X - V),
\end{equation}
from which we obtain
\begin{eqnarray}\label{T2}
\mathbb{T}^{2} = (2V)^{2} \left[\left(\frac{X}{V}\right)^{2}-\frac{X}{V}+1\right].
\end{eqnarray} 
Combining this result with Eq.~\eqref{action}, the action recasts to the K-essence~\cite{armendariz2001essentials} model described by
\begin{equation}\label{Pfunction}
P(X,\phi) =  X-V-\frac{\alpha}{2}M_{\rm p}^{-4(2 \beta-1)} \mathbb{T}^{2\beta}.
\end{equation}
Making use of such a function, we are able to derive the background equations of motion in a spatially flat FLRW spacetime,
\begin{equation}\label{FRW}
ds^2=-dt^2+a^2 \delta_{ij}dx^{i}dx^{j}.
\end{equation}
Here the scale factor $a$ and the field $\phi$ depend only on the cosmic time, \textit{i.e.}, $a=a(t)$ and $\phi=\phi(t)$. Generally, the  corresponding energy-momentum tensor is characterised by the pressure $p=P(X,\phi)$ and the density
\begin{equation}\label{Exphi}
\rho(X,\phi) = 2X P_{,X}(X,\phi) - P(X,\phi),
\end{equation}
where the comma denotes the partial derivative with respect to $X$.
As a result, the background evolution of the scale factor of the universe and the scalar field is given by a set of cosmological equations~\cite{chen2007observational}, i.e.,
\begin{eqnarray}\label{Fridmann}
\rho &=& 3H^{2} M_{\rm p}^2,\\
\dot{\rho} &=& -3H\left(\rho +P \right).
\end{eqnarray} 
where $H=\dot{a}/a$ is the Hubble parameter and the dot stands for derivative with respect to the cosmic time. Hereafter, for the sake of convenience, we fix $M_{\rm p}^2=1$ throughout paper. 
Taking advantage of Eq.~\eqref{Pfunction}, the above relations reduce to 
\begin{equation}\label{Fridmannnew}
3H^{2}= X+V+2\alpha \mathbb{T}^{2(\beta-1)}V^2\Big[(1-4 \beta)\Big(\frac{X}{V}\Big)^2+(2\beta-1)\Big(\frac{X}{V}\Big)+1\Big].
\end{equation}
and
\begin{eqnarray}\label{continuitynew}
\nonumber &\dot{X}&\Big[1-16 \alpha \beta \mathbb{T}^{2 \beta-2}V^{3}\Big(2 (4 \beta-1)\Big(\frac{X}{V}\Big)^3+(1-8 \beta)\Big(\frac{X}{V}\Big)^2+(2 \beta+5) \Big(\frac{X}{V}\Big)-1\Big)\Big]\\
\nonumber &+&\dot{\phi}V'\Big[ 1 + 8 \alpha \beta \mathbb{T}^{2\beta-2}V^{3}\Big((4 \beta-3) \Big(\frac{X}{V}\Big)^3+(11-10 \beta) \Big(\frac{X}{V}\Big)^2-(5-4 \beta) \Big(\frac{X}{V}\Big)+2\Big]\\ &=&  -6 X H \Big[ 1 + 2\alpha \beta \mathbb{T}^{2\beta-2}V\Big(1-2\Big(\frac{X}{V}\Big)\Big],
\end{eqnarray}
where the prime stands for the derivative with respect to $\phi$ and $X=\dot{\phi}^2/2$. We observe that the term $\frac{\partial^2 \mathcal{L}_{\rm m}}{\partial g^{\alpha \beta} \partial g^{\mu\nu}}$ that arises from the variations of the action~\eqref{action} does not contribute to these field equations. It can be verified that this term is identically zero for a canonical scalar field described by $\mathcal{L}_{\rm m}=X-V(\phi)$, see Ref.~\cite{us,Chen:2019dip}. One may also verify that, for $\alpha=0$, as expected, the field equations reduce to their canonical form, specifically the field equations of GR in the presence of a canonical scalar field. 

It's important to note that even when $\alpha \neq 0$, the standard Einstein field equations of GR persist, but now in the presence of a canonical scalar field being complemented by a specific K-essence model determined by the EMSG model under consideration. This aligns with a recent study~\cite{us} suggesting that EMSG, and more broadly, matter-type modified gravity theories like $f(\mathcal{L}_{\rm m})$, $f(g_{\mu\nu} T^{\mu\nu})$, and  $f(T_{\mu\nu} T^{\mu\nu})$, which modify the introduction of the material source in the conventional Einstein-Hilbert (EH) action by incorporating exclusively matter-related terms into the matter Lagrangian density $\mathcal{L}_{\rm m}$, are equivalent to GR. In this equivalence, the usual source is accompanied by a distinct new source, determined by the matter-type modified gravity, which typically interacts non-minimally with the usual source.

\section{Cosmological perturbations and the stability of the model}\label{sec3}
In this section, we attempt to investigate whether the model suffers from the ghost and gradient instabilities. To do this,
we review the analysis of cosmological perturbations  done in Ref.~\citep{chen2007observational} in the comoving gauge. 

The scalar and tensor perturbations of the metric around the background geometry~\eqref{FRW} in the comoving gauge are given by
\begin{equation}
\delta g_{00}=2 A, \hspace{0.5cm} \delta g_{0i}=2 a \partial_{i} B, \hspace{0.5cm} \delta g_{ij}= a^{2} \Big(e^{2 \mathcal{R}} \delta_{ij}+h_{ij}\Big),
\end{equation}
where $A$, $B$, and $\mathcal{R}$ are scalar perturbations, while $h_{ij}$ describe the tensor perturbations. Note that the action~\eqref{action} has $O(3)$ symmetry. Consequently, the scalar and tensor perturbations decouples at the linear order of perturbations. On the other hand, because of the isotropic symmetry, the vector perturbations decay in an expanding Universe. Therefor, we do not consider them here.

After substituting the metric perturbations in the action~\eqref{action}, and expanding it up to the second order and then integrating out
the non-dynamical modes ($A,B$),  one obtains the quadratic action in terms of the dynamical modes ($\mathcal{R}, h_{ij}$)~\cite{chen2007observational,Seery:2005wm} as follows: 
\begin{equation}
S^{(2)}=\frac{1}{2}\int dt d^3 \textbf{x} a^3 \Big[\frac{\varepsilon_{H}}{c_{s}^2}\Big(\dot{\mathcal{R}}^2-\frac{c_{s}^2}{a^2} (\partial \mathcal{R})^2\Big)+\frac{1}{4}\Big((\dot{h}_{ij})^2-\frac{1}{ a^2} (\partial h_{ij})^2 \Big) \Big],
\end{equation}
where speed of sound $c_{s}$ and the standard slow roll parameter $\varepsilon_{H}$ are defined as 
\begin{equation}
c_{s}^2=\frac{P_{,X}}{P_{,X}+2X P_{,XX}}\quad, \quad\quad \varepsilon_{H}\equiv -\frac{\dot{H}}{H^2}=\frac{X P_{,X}}{H^2}.
\end{equation}
For the perturbations to be free from ghost and gradient instabilities, it is necessary that both parameters, $c_{s}^2$ and $\varepsilon_{H}$, are positive. Correspondingly, we require
\begin{equation}
P_{,X}>0, \hspace{1cm} P_{,X}+2 X P_{,XX}>0 \hspace{0.5cm} (\text{or equivalently} \hspace{0.5cm} P_{,XX}>0).
\end{equation}
Exploiting Eq.~\eqref{Pfunction}, these constraints read
\begin{eqnarray}
P_{,X}&=&1+2 \alpha \beta \mathbb{T}^{2 \beta-2} V\Big(1-\frac{X}{V}\Big)\label{PX},\\
P_{,XX}&=&-8 \alpha \beta \mathbb{T}^{2\beta-4} V^{2}\Big[2 (2 \beta-1) \Big(\frac{X}{V}\Big)^2-2(2 \beta-1)\Big(\frac{X}{V}\Big)+(1+\beta)\Big]\label{PXX},
\end{eqnarray}
respectively. Interestingly, for $\beta=1$, one requires that $\alpha<0$ to satisfy the second constraint, whereas the first one confirms $(1+2 \alpha V)>0$. In the next section, we will show the last constraint provides us with $H^2$ in the slow roll approximation. During the inflation epoch under the slow-roll limit where $X \ll V$, Eq.~\eqref{PXX} reduces to
\begin{equation}
P_{,XX}\simeq -2^{2\beta-1} \alpha \beta (1+\beta) V^{2\beta-2}.
\end{equation}
This condition implies that $\alpha<0$  in our scenario for any choice of $\beta$. As a result, in order to satisfy the first condition, the potential must be constrained as follows:
\begin{equation}\label{Vbound1}
V<\frac{1}{2}\Big(|\alpha| \beta\Big)^{\frac{1}{ 1-2\beta}}
\end{equation}
for $\beta>1/2$. In the scale-independent EMSG model with $\beta=1/2$, the first constraint limits us to taking $-2<\alpha<0$. Therefore as far as we confine ourselves to the scalar field inflation models, the cosmological models with $\alpha>0$ in the scale-independent EMSG are ruled out. For example, one of the interesting cosmological solutions presented in~\citep{akarsu4} takes $\alpha=2$\footnote{\cite{akarsu4} uses a parameter $\alpha$ in the action of the theory to identify EMSG corrections. Let us call it $\alpha^*$. The relation between $\alpha^*$ and our $\alpha$ is $\alpha^*=-\alpha/2$.} and reproduces the original steady state universe in the presence of dust~\cite{hoyle}. Based on the stability analysis discussed above, this choice of $\alpha$ is not a healthy one.

A similar challenge arises with the original bouncing EMSG model with $\beta=1$ studied in~\cite{roshan2016energy}. To be specific, after inflation and still deep inside the radiation dominated phase, the Friedmann equation in EMSG is written as~\cite{roshan2016energy}:
\begin{equation}
H^2=\frac{\rho}{3}-\alpha\left(\frac{1}{2}p^2+\frac{4}{3}\rho p+\frac{1}{6}\rho^2\right)
\end{equation}
As $\alpha<0$ due to stability concerns, it is evident that a viable bounce does not exist in this model. Specifically, the first condition for the existence of a bounce, i.e., $H=0$, is not satisfied in the early universe.

As a final remark in this section, it is interesting to note that, in the same spirit, EMSG stabilizes (or destabilizes) a non-relativistic fluid when $\alpha<0$ (or $\alpha>0$). The Jeans analysis in EMSG has been explored in~\cite{kazemi}. EMSG corrections can be combined to define the effective density $\rho_{\text{eff}}$ and pressure $p_{\text{eff}}$ for a fluid with a given fluid density $\rho$ and $p$. The the standard Jeans analysis reveals that the effective sound speed $\mathcal{C}_s$, obtained from $\rho_{\text{eff}}$ and $p_{\text{eff}}$, appears in the dispersion relation of the perturbations instead of the standard sound speed $c_s$. Specifically, the effective sound speed is given by $\mathcal{C}_s^2=c_s^2-\alpha \rho$. Therefore, for $\alpha<0$, EMSG corrections increase the effective sound speed. If we consider the sound speed as the representative of the pressure in the system, acting against gravitational collapse, one may infer that EMSG induces stabilizing effects in the case of $\alpha>0$. For more details, see Ref.~\cite{kazemi}.

 \section{Inflationary solutions}\label{sec4}
 The aim of this section is to obtain a period of inflation in the early
universe by making use of the background equations. Under the slow-roll scheme, when $\dot{\phi}^2 \ll V$ (or equivalently $X \ll V$) and $\ddot{\phi} \ll  H \dot{\phi}$ (or $\dot{X} \ll H X$), Eqs.~\eqref{Fridmannnew} and~\eqref{continuitynew} reduce to
\begin{equation}\label{Fridmannnew1}
3H^{2}\simeq V\left(1+ \alpha \tilde{V}   \right),
\end{equation}
\begin{equation}\label{continuitynew1}
\dot{\phi}V'\left( 1 + 2 \beta \alpha \tilde{V} \right) \simeq -6 X H \left( 1 +\beta \alpha \tilde{V} \right),
\end{equation}
where $\tilde{V}=(2 V)^{\gamma}$ with $\gamma=2 \beta-1$. Note that Eq.~\eqref{Fridmannnew1} demonstrates that there is an upper bound on the potential, i.e., 
\begin{equation}
V <\frac{1}{2} |\alpha|^{\frac{1}{1-2 \beta}} \hspace{0.5cm} \text{for} \hspace{0.5cm}  \beta>\frac{1}{2}
\end{equation}
This is situated within the acceptable bound~\eqref{Vbound1}, necessary to evade the ghost instability. In addition, in the particular case $\beta=1/2$, the positiveness of the energy density modifies the lower bound on $\alpha$, namely $-1<\alpha<0$.
To sum up, to remedy the ghost and gradient instabilities, the coupling constant $\alpha$ must be placed at
  \begin{equation}
\begin{matrix}
  \alpha<0 \hspace{1.5cm} \text{for} \hspace{1cm} \beta>\frac{1}{2},\\
  -1<\alpha<0 \hspace{0.5cm}\text{for} \hspace{1cm} \beta=\frac{1}{2}.
  \end{matrix}
  \end{equation}
It is worth noting that for the EMPG model with $\beta=1$, $\alpha>0$ is not well-behaved when the growth of the linear matter perturbations is concerned~\cite{farsi}. Furthermore, this finding raises severe criticism about the inflationary solutions obtained in~\cite{faraji}, where the authors assume $\alpha>0$ without considering the stability of the model. 

Taking the time derivative of the both sides of Eq.~\eqref{Fridmannnew1} and using it together with Eq.~\eqref{Fridmannnew1}, we obtain the Hubble slow-roll parameter
\begin{equation}\label{epsilonv1}
\varepsilon_{H} \simeq -\frac{1}{2 \gamma}\Big(\frac{\tilde{V} '}{\tilde{V} }\Big)\Big(\frac{\dot{\phi}}{H}\Big)\Big(\frac{1+2\alpha \beta \tilde{V} }{1+\alpha \tilde{V} }\Big).
\end{equation}
 From Eq.~\eqref{continuitynew1}, it is straightforward to write the ratio $\dot{\phi}/H$ as 
\begin{equation}\label{phidoth}
\frac{\dot{\phi}}{H} = -\Big(\frac{\tilde{V} '}{\gamma \tilde{V} }\Big)\frac{1+2 \alpha \beta\tilde{V} }{\Big(1+\alpha \beta
 \tilde{V} \Big)\Big(1+\alpha \tilde{V} \Big)}.
\end{equation}
Thus the slow roll parameter~\eqref{epsilonv1} converts to
\begin{equation}\label{epsilonvnew}
\varepsilon_{H} = \frac{1}{2 \gamma^2}\bigg(\frac{\tilde{V} '}{\tilde{V} }\bigg)^{2}\frac{\Big(1+2\alpha \beta \tilde{V} \Big)^{2}}{\Big(1+\alpha \beta \tilde{V} \Big)\Big(1+\alpha  \tilde{V} \Big)^{2}}.
\end{equation}
The other Hubble slow-roll parameter $\eta_{H}$ can be expressed as
\begin{equation}
\begin{split}
\eta_{H}=\frac{\dot{\varepsilon_{H}}}{H \varepsilon_{H}}=\frac{\varepsilon_{H}'}{\varepsilon_{H}} \frac{\dot{\phi}}{H}=&\frac{1}{\gamma}
 \bigg(\frac{\tilde{V}'}{\tilde{V}}\bigg)^{2} \frac{2+\alpha \tilde{V}\Big(4+3\beta+\alpha \beta \tilde{V} (9+2 \beta +6 \alpha \beta \tilde{V})\Big)}{\Big(1+\alpha \tilde{V}\Big)^{2}\Big(1+\alpha \beta  \tilde{V}\Big)^{2}}\\&
-\frac{2}{\gamma} \bigg(\frac{\tilde{V}''}{\tilde{V}}\bigg) \frac{\Big(1+2 \alpha \beta \tilde{V}\Big)}{\Big(1+\alpha \tilde{V}\Big)\Big(1+\alpha \beta  \tilde{V}\Big)}.
\end{split}
\label{etavnew}
\end{equation}
At the limit $\alpha \to 0$, both slow roll parameters reduce to the standard form. Great care must be taken when investigating the slow roll parameters. First, during the inflation era these parameters must be much smaller than one, namely $\varepsilon_{H}\ll 1$ and $\eta_{H}\ll 1$. Second, their evolution must take at least 50-60
number of e-folds to solve the flatness and the horizon problems. At the end, inflation ends
when either of the slow-roll parameters tends to unity. Moreover, in the slow-roll limit, the sound speed can be estimated as
 \begin{equation}\label{speed1}
 c_{s} \simeq 1+\frac{\alpha \beta (1+\beta) (1+\alpha \tilde{V}) \tilde{V}}{6 (1+\alpha \beta \tilde{V})} \Big(\frac{\dot{\phi}}{H}\Big)^2.
 \end{equation} 
Since $\beta>0$ and $1+\alpha \tilde{V}>0$, it is clear that the case of $\alpha>0$ leads to superluminal speed of sound which naturally implies the violation of Null Energy Condition (NEC) in regular bounce models. It should be stressed that the superluminal sound speed does not necessarily violate the causality~\cite{mukhanov}. We leave this as a subject of study for future works.

Furthermore, using Eq.~\eqref{phidoth}, one can express the above relation as function of the potential and its derivatives.
Following~\cite{chen2007observational,Seery:2005wm}, the scalar and tensor power spectrum\footnote{An exact determination of $\mathcal{P}_{\mathcal{R}}$ is provided by solving the Mukhanov-Sasaki equation given by 
\begin{equation}\label{SMeq}
\frac{d^2 v_{k}}{d \tau^2}+\Big(c_{s}^2 k^2-\frac{1}{z} \frac{d^2 z}{d \tau^2} \Big)v_{k}=0
\end{equation}
where $\tau$ is the conformal time, the variable $k$ relates to the comoving scale by $\lambda=2 \pi/k$, subscript $k$ indicates the momentum space~\cite{garriga1999perturbations}.  Additionally, the curvature perturbation $\mathcal{R}$ is related to $v$ via $\mathcal{R}=v/z$ and the background variable $z$ is defined as $z=\frac{a^2(\rho+P)}{c_{s}^2 k^2}$. Finally, the power spectrum of the scalar mode at the horizon crossing where $aH=c_{s}k$ can be expressed as
\begin{equation}\label{powerN}
\mathcal{P}_{\mathcal{R}}=\frac{k^3}{2 \pi}|\frac{v_{k}}{z}|^2
\end{equation}

 } in the slow roll approximation are also given by
\begin{equation}\label{slopower}
\mathcal{P}_{\mathcal{R}} \simeq \frac{1}{8 \pi^2} \frac{H^2}{\varepsilon_{H} c_{s}}|_{c_{s}k=aH}, \hspace{1cm}\mathcal{P}_{h}=\frac{2}{\pi^2} H^2|_{k=aH}
\end{equation}
Having calculated power spectra, we can also calculate the
spectral index $n_s$ and the tensor-to-scalar ratio $r$ as follows:
\begin{equation}\label{nsanalatic}
n_{s}-1\equiv \frac{d \ln \mathcal{P}_{\mathcal{R}}}{d \ln k}\simeq -2 \varepsilon_{H}-\eta_{H}-s\quad,
\end{equation}
\begin{equation}\label{rformula}
r\equiv \frac{\mathcal{P}_{h}}{\mathcal{P}_{\mathcal{R}}}=16 \varepsilon_{H} c_{s}\quad,
\end{equation}
where $s\equiv \dot{c_s}/(H c_{s})$. Moreover, the tensor spectral index $n_{T}$ is given by
\begin{equation}
n_{T}\equiv \frac{d \ln \mathcal{P}_{h}}{d \ln k}=-2 \varepsilon_{H}
\end{equation}
and it satisfies the consistency relation $r=-8 c_{s} n_{T}$. The parameters $r$ and $n_{s}$ can be limited by current observational constraints on inflationary parameters~\cite{Planck:2018jri,BICEP2:2018kqh}. Using Eqs.~\eqref{epsilonvnew} and~\eqref{etavnew}, one can express the above parameter in terms of the scalar potential $V(\phi)$ and its derivatives, rather than with $\phi$ itself.

\subsection{Predictions of the model with chaotic potentials}
Up to now, we have obtained the equations for the slow roll parameters, the spectral index, and tensor to scalar ratio without selecting a particular potential type. In this subsection, we proceed with using the chaotic potentials $V(\phi)=\frac{A}{M_{\rm p}^{n-4}} \phi^{n}$, where $A$ is a dimensionless coefficient. For such a potential, $n$ takes rational numbers and $A$ stands for the normalisation parameter given by the amplitude of the scalar power spectrum at the CMB pivot scale.

The number of e-folding ($N$) measures how much inflation took place to the end of inflation and is defined as 
\begin{equation}
N=-\int_{t_{\rm e}}^{t} H dt=-\int_{\phi_{\rm e}}^{\phi} \frac{H}{\dot{\phi}}d\phi=\int_{\tilde{V}_{\rm e}}^{\tilde{V}} \frac{1}{2 \gamma \varepsilon_{H}} \frac{d\tilde{V}}{\tilde{V}} \frac{1+2\alpha \beta  \tilde{V}}{1+\alpha \tilde{V}}\quad,
\end{equation}
where we have used Eq.~\eqref{epsilonv1}. Here the subscript $\rm e$ indicates the value of the quantities at the end of the inflation. By substituting $\tilde{V}=\tilde{A} \phi^{\gamma n}$, where $\tilde{A}=(2 A)^{\gamma}$ into the above expression, it is straightforward to verify that
\begin{equation}\label{Nfunction}
N=\frac{1}{2 n}\Big(\frac{\tilde{V}}{\tilde{A}}\Big)^{\frac{2}{\gamma n}}\Big[1+\frac{ \alpha \tilde{V}}{2+\gamma n} \Big(1- \gamma {}_{2}F_{1}(1,1+\frac{2}{\gamma n},2+\frac{2}{\gamma n},-2 \alpha \beta  \tilde{V})\Big)  \Big] .
\end{equation}
This concurs well with~\cite{Li:2012vta,Unnikrishnan:2012zu} in the $\alpha\to 0$ limit. Notice that the potential $\tilde{V}$ in~\eqref{Nfunction} should be calculated at the beginning of the inflation. Moreover, for the scale-independent EMSG ($\beta=1/2$, or $\gamma \to 0$), the above general relation reduces to
\begin{equation}
N=\frac{2+\alpha}{4n}\Big(\frac{V}{A}\Big)^{\frac{2}{n}} \hspace{0.5cm} \text{and conversely} \hspace{0.5cm} V=A \Big(\frac{4n N}{2+\alpha}\Big)^{\frac{n}{2}}.
\end{equation}
While for the case $\gamma>0$, due to the existence of the hyper-geometrical function in Eq.~\eqref{Nfunction}, it becomes difficult to express the potential function $V$ as a function of $N$ in reverse. However, it can be done in the situation that $|\alpha|^{1/\gamma} V \leq \mathcal{O}(\sqrt{\varepsilon_{H}^{SC}})<1/2$ during inflation.\footnote{The symbol $\rm SC$ stands for the standard chaotic inflation.}  By making this assumption, and defining the expansion parameters $\epsilon=\alpha A$ and $\tilde{\epsilon}=\alpha \tilde{A}$, one obtains
 \begin{eqnarray}
 \label{VAexpand}
\frac{\tilde{V}}{\tilde{A}}\simeq \Big(2n N\Big)^{\frac{\gamma n}{2}}\Big[1+f_1(N) \tilde{\epsilon}+f_2(N) \tilde{\epsilon}^2+\mathcal{O}(\tilde{\epsilon}^3)\Big] ,
 \end{eqnarray}
 where the auxiliary functions $f_1$ and $f_2$ are defined as
 \begin{equation}
 \begin{split}
&f_1(N)=\gamma n \frac{(2n N)^{\frac{\gamma n}{2}}(\beta-1)}{2+\gamma n},\\&
 f_2(N)=\gamma n \frac{(2nN)^{\gamma n}}{2(1+\gamma n)(2+\gamma n)^2} f(\beta),
 \end{split}
 \end{equation}
 and the function $f(\beta)$ is given by
 \begin{equation}
 f(\beta)=2-6 \beta^2+n^2 \gamma^2\Big(3+(\beta-5)\beta\Big)+n \gamma \Big(5-3 \beta (2+\beta)\Big).
 \end{equation}
Although we have considered the above relation up to the second order, as long as $|\alpha|^{1/\gamma} V$ is close to the bound $\sqrt{\varepsilon_{H}^{SC}} \sim \mathcal{O}(10^{-1})$, one needs to take higher orders, at least the forth order, to make an excellent agreement between analytical and numerical results as presented in Fig. \ref{FigAN}.

 Now, by combing Eqs.~\eqref{epsilonvnew},~\eqref{etavnew} with Eq.~\eqref{VAexpand}, we can obtain the relation between the
 slow-roll parameters and $N$ approximated to the second order of $\alpha \tilde A$
 \begin{equation}\label{slowrollN}
\varepsilon_{H} \simeq  \frac{n}{4 N}\Big[1+\frac{f_1(N)(2+n (3 \beta-2))}{n (\beta-1)}  \tilde{\epsilon}
+ \frac{2f_2(N)}{\gamma n f(\beta)}(g(\beta)-f(\beta)) \tilde{\epsilon}^2
 +\mathcal{O}(\tilde{\epsilon}^3)\Big],
 \end{equation}
 \begin{equation}
\eta_{H}\simeq  \frac{1}{ N}\Big[1-\frac{\gamma f_1(N)(2+n (3 \beta-2))}{2(\beta-1)} \tilde{\epsilon}
+ \frac{f_2(N)}{n\gamma f(\beta)} (p(\beta)-2 f(\beta)) \tilde{\epsilon}^2
 +\mathcal{O}(\tilde{\epsilon}^3)\Big],
 \end{equation}
 where the functions $g(\beta)$ and $p(\beta)$ are defined as
 \begin{equation}
 g(\beta)=(1+n \gamma)(2+n \gamma) \Big(3+5 n \gamma+\beta (n\gamma(4 \beta-11)-(4+3 \beta))\Big),
 \end{equation}
 \begin{equation}
 p(\beta)=(1+\gamma n)(2+\gamma n)(2-6 \beta^2 + 2 n \beta \gamma (7 \beta-3)+n^2\gamma^2 (\beta(10+\beta)-6)).
 \end{equation}
 Note that our finding confirms the relation $\eta_{H} =-d\ln \epsilon_{H}/dN$. In addition, the sound speed $c_{s}$ is obtained to be
 \begin{equation}\label{csexpand}
 c_{s}\simeq 1+\frac{2n}{3 N} (2 n N)^{\frac{n \gamma}{2}} \beta (\beta+1)\tilde{A} \alpha \Big[1+ \frac{2n \gamma(2nN)^{\frac{n \gamma}{2}}(\beta^2-1)}{(2+n\gamma)(1+\beta)}\tilde{\epsilon}+ \mathcal{O}(\tilde{\epsilon}^{2}) \Big].
 \end{equation}
This allows to calculate the slow-roll parameter $s\equiv \dot{c_s}/(H c_{s}) = -d(\ln c_{s})/dN$.  
 Therefore, a formal solution for the spectral index~\eqref{nsanalatic} is given by
 \begin{eqnarray}
\nonumber   n_{s}-1 &\simeq & -\frac{1}{N}\Big[\frac{2+n}{2}+\Big\{\frac{(1-\gamma)f_1(N)(2+n(3 \beta-2))}{2(\beta-1)}-\frac{n}{24 N^2} (2 n N)^{\frac{n \gamma}{2}} \beta (\beta+1)(n \gamma-2)\Big\} \tilde{\epsilon}\quad,\\
\nonumber &+& \Big\{\frac{n^2}{288 (2+n \gamma) N^2} (2 n N)^{n \gamma} \beta (\beta+1) \Big(-48  \gamma N (\beta-1) (n \gamma-1)+\beta (\beta+1) (n^2 \gamma^2-4)\Big)\\
&+& \frac{f_2(N)}{n \gamma f(\beta)} \Big( p(\beta)+n g(\beta)-(2+n)f(\beta)\Big)\Big\} \tilde{\epsilon}^2
 + \mathcal{O}(\tilde{\epsilon}^3)\Big].
 \end{eqnarray}
It should be noted that all the parameters mentioned above can be expressed up to arbitrary order of $\alpha \tilde{A}$. Here, for the sake of simplicity, we have written the first three leading terms. As we can see, the result for the spectral index for the case $\alpha=0$ in~\cite{Li:2012vta,Unnikrishnan:2012zu} is modified by the orders of $\alpha \tilde{A}$ in our scenario. Note that it is easy to write down $r$ in the terms of $N$ by combining Eqs.~\eqref{rformula},~\eqref{slowrollN}, and~\eqref{csexpand} together.  

For the special case of the scale-independent EMSG ($\beta=1/2$, or $\gamma \to 0$), and by taking the potential $V=A \phi^n$, we arrive at 
\begin{eqnarray}
c_{s}&= & 1+\frac{n\alpha (\alpha+1)}{4 N (2+ \alpha)^2}\label{cs1}\quad,\\
\varepsilon_{H} &=& \frac{n}{4 N}, \hspace{0.5cm} \eta_H=\frac{1}{N}\quad,\\
\label{rbeta12}r &=& 16 \varepsilon_{H} c_{s}=\frac{4n}{N}+\frac{n^2\alpha (\alpha+1)}{ N^2 (2+ \alpha)^2}\simeq \frac{4n}{N}\quad,\\
\label{nsbeta12}n_{s}&-& 1 =  -\frac{2+n}{2 N}.
\end{eqnarray}
As a consequence of the above discussion, the scale-independent EMSG predicts the same inflationary parameters as those predicted by the standard single-field power-law inflation~\cite{Li:2012vta,maldacena2003non}.

Now, let us consider another special EMSG model with $\beta = 1$ (or $\gamma \to 1$). By using the chaotic potential $A \phi^2$, the parameters for this model can be expressed up to fifth order of $\epsilon$ as follows.
\begin{eqnarray}
 c_{s} &\simeq& 1+\frac{8 \epsilon}{3}- \frac{1024}{3} N^2 \epsilon^3+ \frac{16384}{3} N^3 \epsilon^4 + \mathcal{O}(\epsilon^5)\quad,\\
\nonumber  r &\simeq & \frac{8}{N}+64\Big(1+\frac{1}{3 N}\Big)  \epsilon + \Big(\frac{512}{3}-\frac{2560}{3} N\Big) \epsilon^2+ \frac{2048}{9} N (27 N-22) \epsilon^3\\
 &+&\frac{16384}{9} N^2 (\frac{22}{5} N+21) \epsilon^4+ \mathcal{O}(\epsilon^5)\quad,\\
\nonumber  n_{s}&-&1 \simeq - \frac{2}{N}-\frac{512}{3} N \epsilon^2+ \frac{512}{3} N (-4+27 N) \epsilon^{3}\\
&-&\frac{16384}{9} N \Big(N(\frac{208}{5} N-9) -1\Big) \epsilon^{4}
 + \mathcal{O}(\epsilon^5).
 \end{eqnarray}
It is obvious that the EMSG corrections provides us with a clear improvement on the value of $r$ and $n_{s}$ for the standard single field inflation. Specifically, due to the negative nature of $\alpha$ and $\epsilon$, all corrections have the potential to decrease $r$ and $n_{s}$ in comparison to standard single field inflation, bringing them closer to current observational constraints. For instance, in order to be compatible with large scale CMB observations~\cite{Planck:2018jri,BICEP2:2018kqh} at the pivot scale $k_{*}=0.05 \,\text{Mpc}^{-1}$ with
\begin{equation}\label{Planckcon}
0.956<n_{s}<0.978,\quad r(k_{*})\leq 0.06 \quad \text{at} \quad \text{95\%\,C.L.},
\end{equation} 
which in turn implies for $\epsilon=\alpha A$ that
\begin{equation}\label{bound1}
-0.00064 < \alpha A \leq -0.00052
\end{equation}
for $N=70$, whereas for $N=50$ and $N=60$ the EMSG model cannot put into the observational constraint. In Fig. \ref{FigAN}, the solid curves present the values of $\{n_{s},r\}$ with respect to $\epsilon$ by using analytical relations for the power law potential $V=A \phi^n$ with $n=2$ and $n=3/2$, respectively. Moreover, we compared these results with numerics that are depicted by colored points in Fig. \ref{FigAN}. More importantly, there is satisfactory agreement between numerics and analytics. 
In the next subsection, we explain our numerical method.   
\begin{figure}
\includegraphics[scale=0.49]{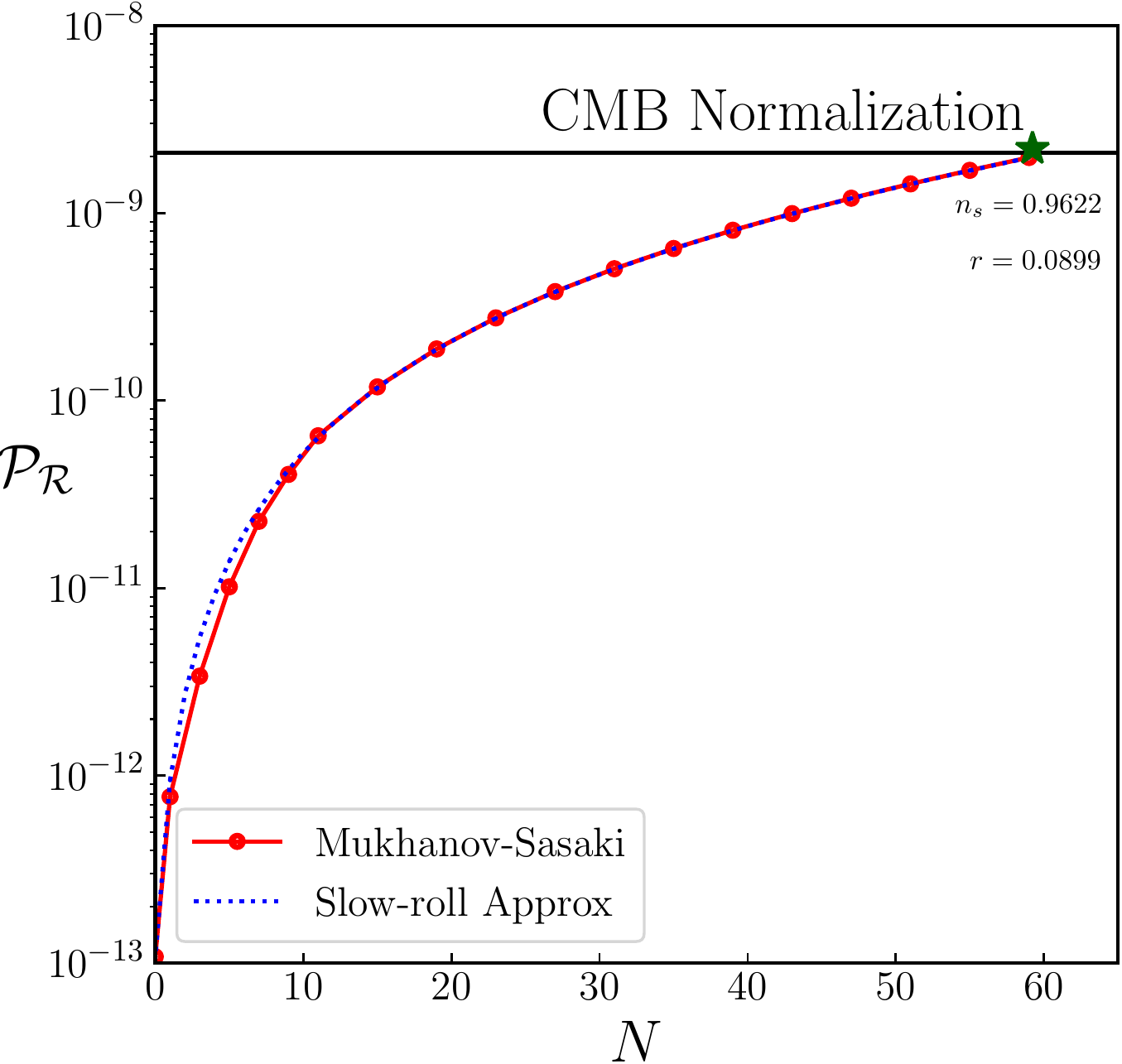}\hspace{0.5cm}
 \includegraphics[scale=0.49]{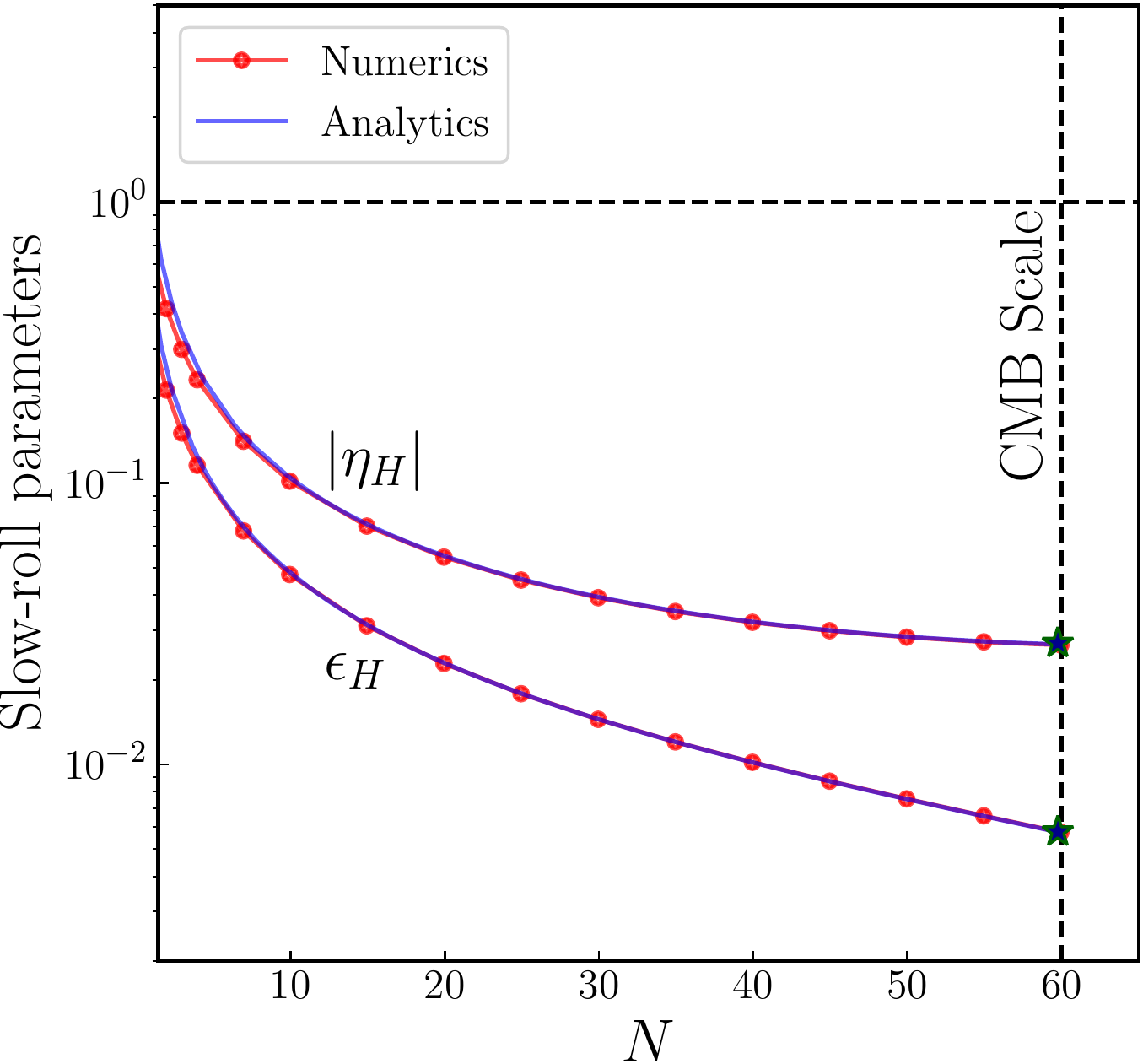}
 \caption{  \textit{Left panel}: The scalar power spectrum $\mathcal{P}_{\mathcal{R}}$ is plotted as a function of the number of e-folds before the end of inflation $N_{\rm e}$ (a) by using the slow-roll approximation~\eqref{slopower} (dotted blue) and (b) by numerically solving the Mukhanov- Sasaki equation~\eqref{powerN} (red dots). Interestingly, both methods give the same results for a smoothly varying potential, in which case $\mathcal{P}_{\mathcal{R}}$ increases monotonically with increasing $N_{\rm e}$.
 \textit{Right Panel}: The evolution of the slow-roll parameters during the inflationary phase (a) by numerical solving background equations~\eqref{Fridmannnew} and~\eqref{continuitynew} (b) by using Eqs.~\eqref{epsilonvnew},~\eqref{etavnew}, and~\eqref{Nfunction}  under slow roll approximation. Note that the slow roll conditions $\varepsilon_{H},\eta_{H}\ll1$ remain satisfied during inflation. 
  Here we have considered the  potential $V=A \phi^2$ in which $\alpha A=-0.00046$ for EMSG model with $\beta=1$ in both panels. A similar behavior is held for the other power law potential types. \label{Fig1}}
\end{figure}
\begin{figure}
\centering
 \includegraphics[scale=0.4]{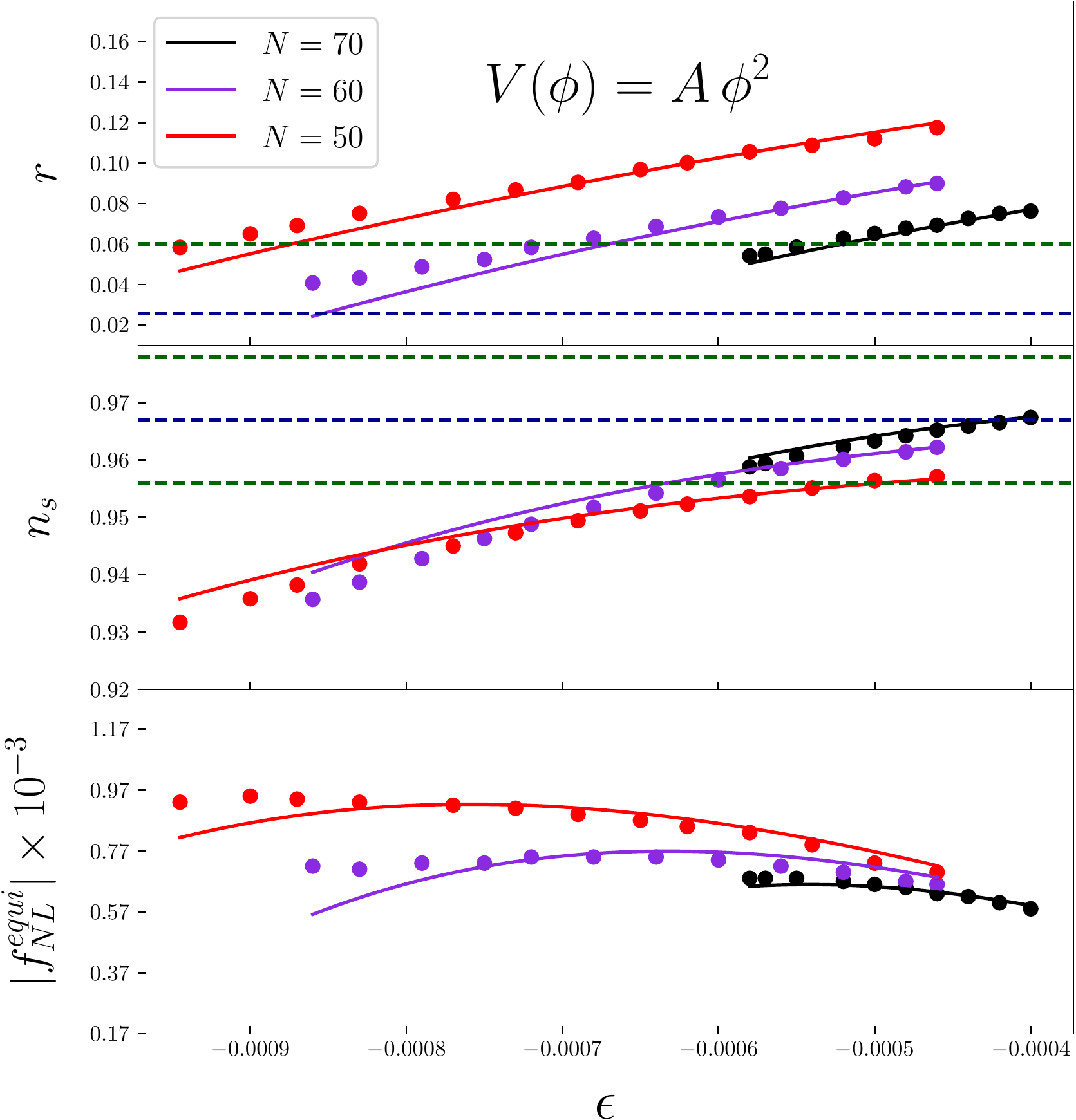}\hspace{0.5cm}
 \includegraphics[scale=0.4]{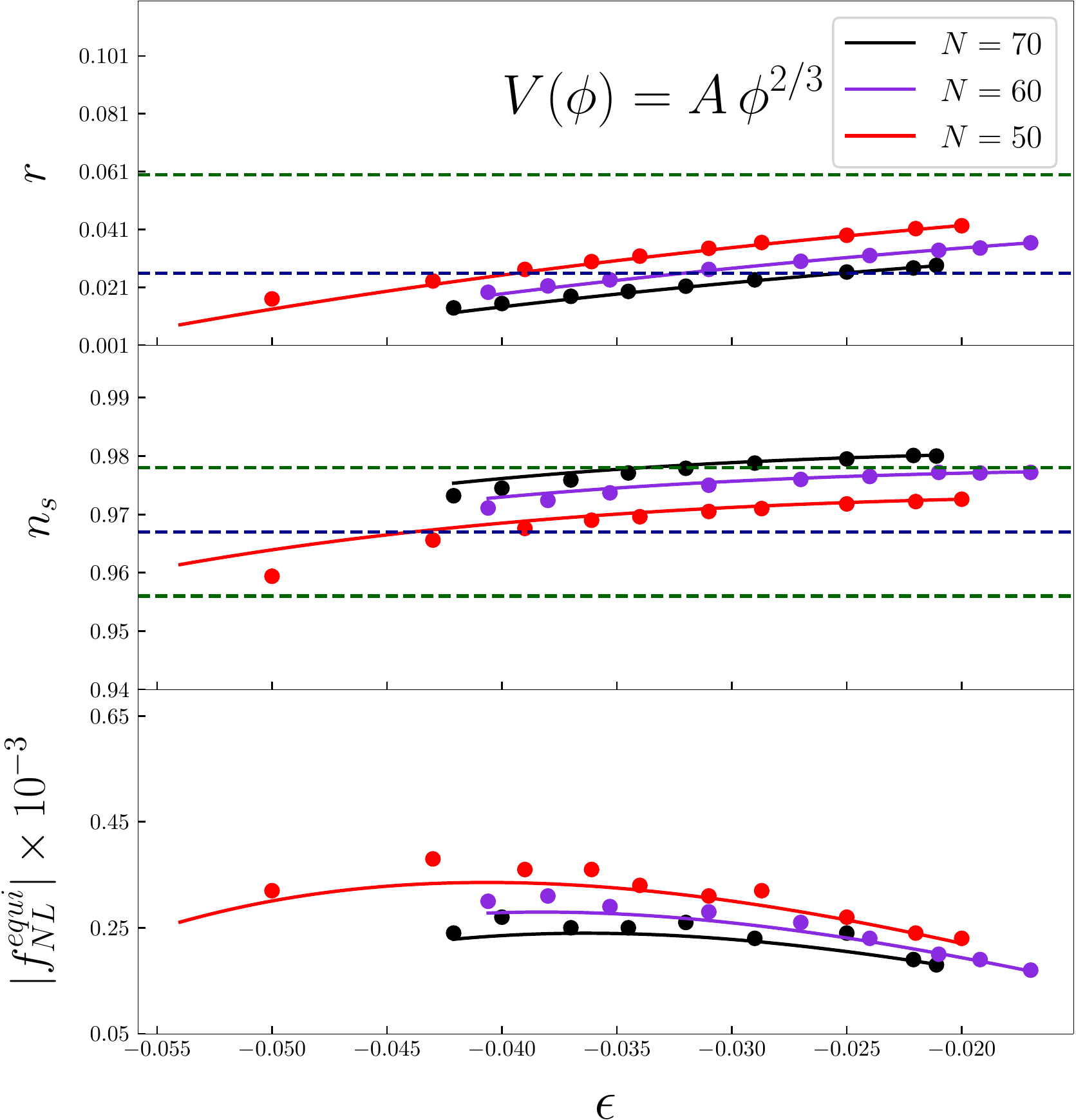}
 \caption{ A comparison between analytical and numerical results for inflationary parameters in EMSG model with $\beta=1$ by considering a chaotic potential case, \textit{left panel}: $V=A \phi^2$ and \textit{right panel}: $V=A \phi^{2/3}$. Note that the bullet points represent numerical data, while the solid curves depict analytical results. Additionally, the horizontal dashed lines in different colors indicate the relevant observational measurements: BK15 (green) and BK18 (blue)\label{FigAN}.}
\end{figure}

Before proceeding, let us analytically investigate another crucial inflationary parameter, namely the non-Gaussianity parameter $f_{\rm{NL}}$ in our scenario. Observational limitations on primordial non-Gaussianities have the potential to constrain the parameter space of our model. In fact, according to the Planck observations,  the amplitude of the non-Gaussianity parameter $f_{\rm{NL}}$ in the equilateral shape is constrained to be $f^{\rm{equi}}_{\rm{NL}}=-26\pm 47$ ($68 \%$ C.L.)~\cite{Planck:2018jri}.
 In the K-essence models, the non-Gaussianity parameter in the equilateral limit at leading order is given by
\begin{equation}\label{fNl1}
f^{\rm{equi}}_{\rm{NL}}\simeq \frac{5}{81} \Big(\frac{1}{c_{s}^{2}}-1-2 \frac{\lambda}{\Sigma}\Big)-\frac{35}{108} \Big(\frac{1}{c_{s}^{2}}-1\Big),
\end{equation}
where
\begin{eqnarray}\label{lamda}
\lambda=X^2 P_{,XX}+\frac{2}{3}X^{3} P_{,XXX} \quad,\quad
\Sigma = X P_{,X}+2 X^{2}P_{,XX}.
\end{eqnarray}
In the slow-roll approximation, one finds, from Eq.~\eqref{speed1} that 
\begin{equation}
2\frac{\lambda}{\Sigma} \simeq -\frac{\alpha \beta (\beta+1)(1+\alpha \tilde{V}) \tilde{V}}{3(1+\alpha \beta \tilde{V})} \Big(\frac{\dot{\phi}}{H}\Big)^2\simeq \frac{1}{c_{s}^2}-1.
\end{equation}
By substituting the above relation into~\eqref{fNl1}, we obtain
\begin{equation}
f^{\rm{equi}}_{\rm{NL}}\simeq -\frac{35}{108} \Big(\frac{1}{c_{s}^{2}}-1\Big).
\end{equation}
Obviously, the leading order contribution in the non-Gaussianity parameter vanishes similar to the DBI models (with a non-canonical kinetic term) where large non-Gaussianity of the equilateral type can be generated~\cite{alishahiha2004dbi}. 

For the EMSG model with $\beta=1$, in the valid range of the EMSG parameter $\epsilon$~\eqref{bound1}, the non-Gaussianity parameter for a chaotic potential with $n=2$ can be estimated to be
\begin{equation}
f_{\rm{NL}}^{\rm{equi}} \simeq \frac{70}{81} \epsilon \sim \mathcal{O}(10^{-3}) \sim \mathcal{O}(\varepsilon_{H}^{SC}).
\end{equation} 
This is consistent with the numerical results, as shown in Figure \ref{FigAN}. 
Clearly, the non-Gaussianity parameter $f_{\rm{NL}}^{\rm{equi}}$ falls within an acceptable range, but its value is so small similar to the standard single field inflation~\cite{maldacena2003non}. Consequently, for the models which are built of the energy momentum tensor associated with the canonical matter Lagrangian, the non-Gaussianity parameter $f_{\rm{NL}}^{\rm{equi}}$ gets very small values. However, we expect that a non-canonical matter Lagrangian leads to a large equilateral non-Gaussianity in our scenario. Several studies, for example~\cite{unnikrishnan2013resurrecting,alishahiha2004dbi,Seery:2005wm,chen2007observational} have shown that a suitable choice of non-canonical matter Lagrangian in K-inflation model can generate large $f_{\rm{NL}}^{\rm{equi}}$ by altering the speed of sound. We will investigate such models in the next section. Before moving on, let us now compare analytical results  with numerics.

\subsection{Numerics}
To obtain numerical values for $n_{s}$, $r$, and $f_{\rm{NL}}^{\rm{equi}}$, it is necessary to first obtain numerical solutions for $\phi$ and $H$, as required by the background equations, Eqs.~\eqref{Fridmannnew} and~\eqref{continuitynew}, as a function of the e-folding number $N$. In our numerical calculations, the variable $N$ represents the number of e-foldings before the end of inflation. Thus, $N=N_{i}$ denotes the beginning of inflation, while $N=N_{\rm e}=0$ corresponds to the end of inflation. We also define $N^{*}$ as the number of e-foldings before the end of inflation when the CMB pivot scale $k^{*}=0.05, \text{Mpc}^{-1}$ exited the comoving Hubble radius. For convenience, we consider three values of $N^{*}$, namely $N^{*}=\{50,60,70\}$.

It is common for power law potentials of the form $V= A \phi^{n}$ to predict a smooth scalar power spectrum $\mathcal{P}_{\mathcal R}$ that decreases monotonically from the largest scales ($N \sim N^{*}$) to the smallest scales ($N \sim 0$), as shown in the left panel of Fig.~\ref{Fig1}.It is worth noting that the value of the coefficient $A$ is fixed by the CMB normalization on the power spectrum, which requires $\mathcal{P}_{R}\simeq 2.1 \times 10^{-9}$ at the CMB pivot scale $k^{*}$~\cite{Planck:2018jri}. Additionally, we show the evolution of the slow-roll parameters $\epsilon_{H}$ and $\eta_{H}$ in the right panel of Fig.~\ref{Fig1}. As can be seen, there is perfect agreement between the numerical results (without the slow-roll approximation) and those obtained using the slow-roll approximation. To illustrate this, we plot the slow-roll parameters~\eqref{epsilonvnew} and~\eqref{etavnew} as a function of $N$ from Eq.~\eqref{Nfunction}, by smoothly varying the potential form from $V_{\rm e}=A\phi_{\rm e}^n$ to $V_{i}=A\phi_{i}^n$. It should be noted that $V_{\rm e}$ (or equivalently $\phi_{\rm e}$) is determined when the slow-roll parameter $\varepsilon_{H}$ reaches unity, i.e., when the inflationary phase ends.

\begin{figure}
\centering
 \includegraphics[scale=0.45]{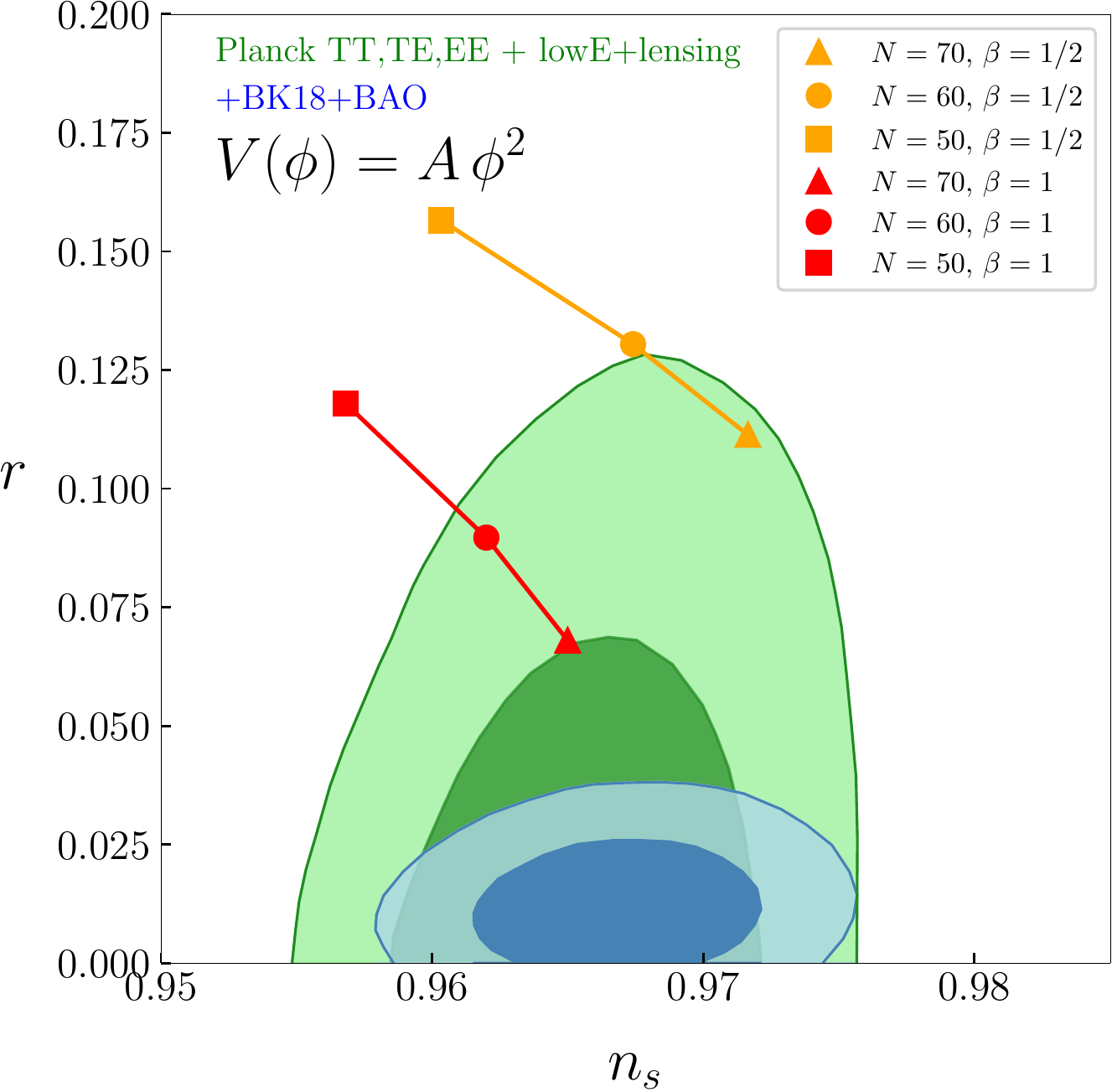}\hspace{0.5cm}
 \includegraphics[scale=0.45]{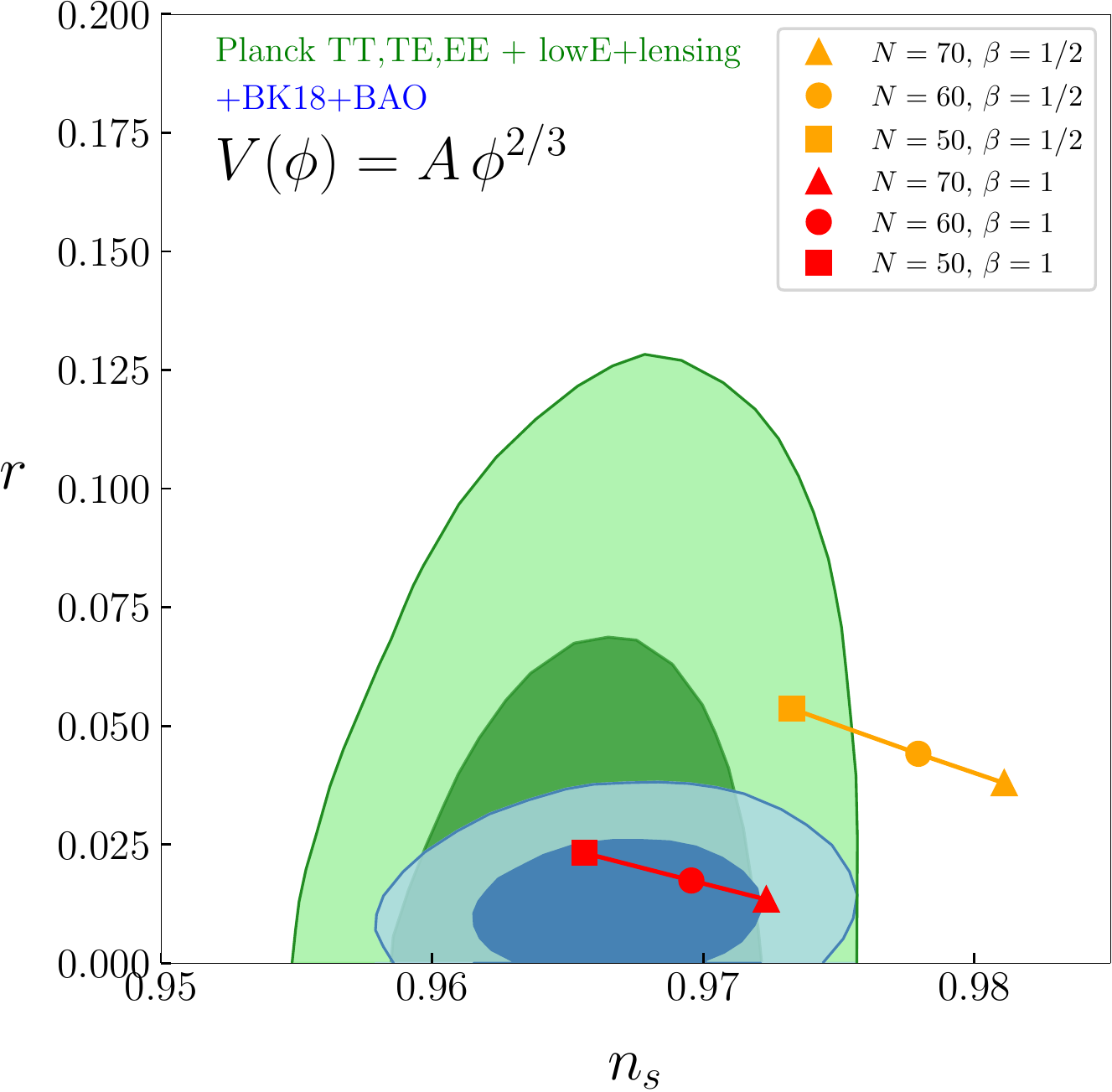}
 \caption{ Tensor-to-scalar ratio vs spectral index for EMSG model with the power law potential $V=A \phi^{n}$ for  $n=2$ (right) $n=2/3$ (left), compared to the data of Ref.~\cite{ade2021improved}.  \label{Fig2}}
\end{figure}

In Fig. \ref{Fig2}, we also present the tensor-to-scalar ratio as a function of the spectral index, along with the observational constraints from the \textit{Planck} 2018 data, as well as BICEP/\textit{Keck} (BK15~\cite{aghanim2020planck} and BK18~\cite{ade2021improved}) data and BAO data\footnote{The BK18 analysis yielded a $95 \%$ confidence constraint from BK15 as $r_{0.05}<0.07$ improved to $r_{0.05}<0.036$. Additionally, the BK18 simulations result in a median 95 upper limit of $r_{0.05} < 0.019$.}. Fig. \ref{Fig2} is interesting for several reasons. Firstly, the numerical values of $r$ and $n_{s}$ for the case of $\beta=1/2$ are located exactly where those predicted by the standard single-field inflation for various values of $N$~\cite{aghanim2020planck}. This finding confirms the analytical relations~\eqref{rbeta12} and~\eqref{nsbeta12} in the slow-roll regime. Secondly, incorporating the correction of the energy-momentum tensor with $\beta=1$ into the standard canonical scalar field inflation improves the predicted values of $\{r,n_{s}\}$, bringing them into agreement with recent observational constraints. For instance, for the potential $V=A \phi^{2/3}$, the predicted values fall entirely within the region determined by the BK18 results~\cite{ade2021improved}.   
 
Furthermore, in Fig. \ref{FigAN}, we compare the numerical and analytical results for $\{n_{s},r,f_{\rm{NL}}^{\rm{equi}}\}$, revealing a noteworthy finding: the numerical values of $\{n_{s},r,f_{\rm{NL}}^{\rm{equi}}\}$ are in close agreement with the analytical predictions discussed earlier. Additionally, we observe that $f^{\rm equil}_{\rm{NL}}$ yields small values, similar to those of standard chaotic inflation. However, to address this issue, we consider a non-canonical Lagrangian in our scenario. 

\section{Non-canonical matter Lagrangian and large non-gaussianities} \label{sec5} 
\begin{figure}
\centering
 \includegraphics[scale=0.45]{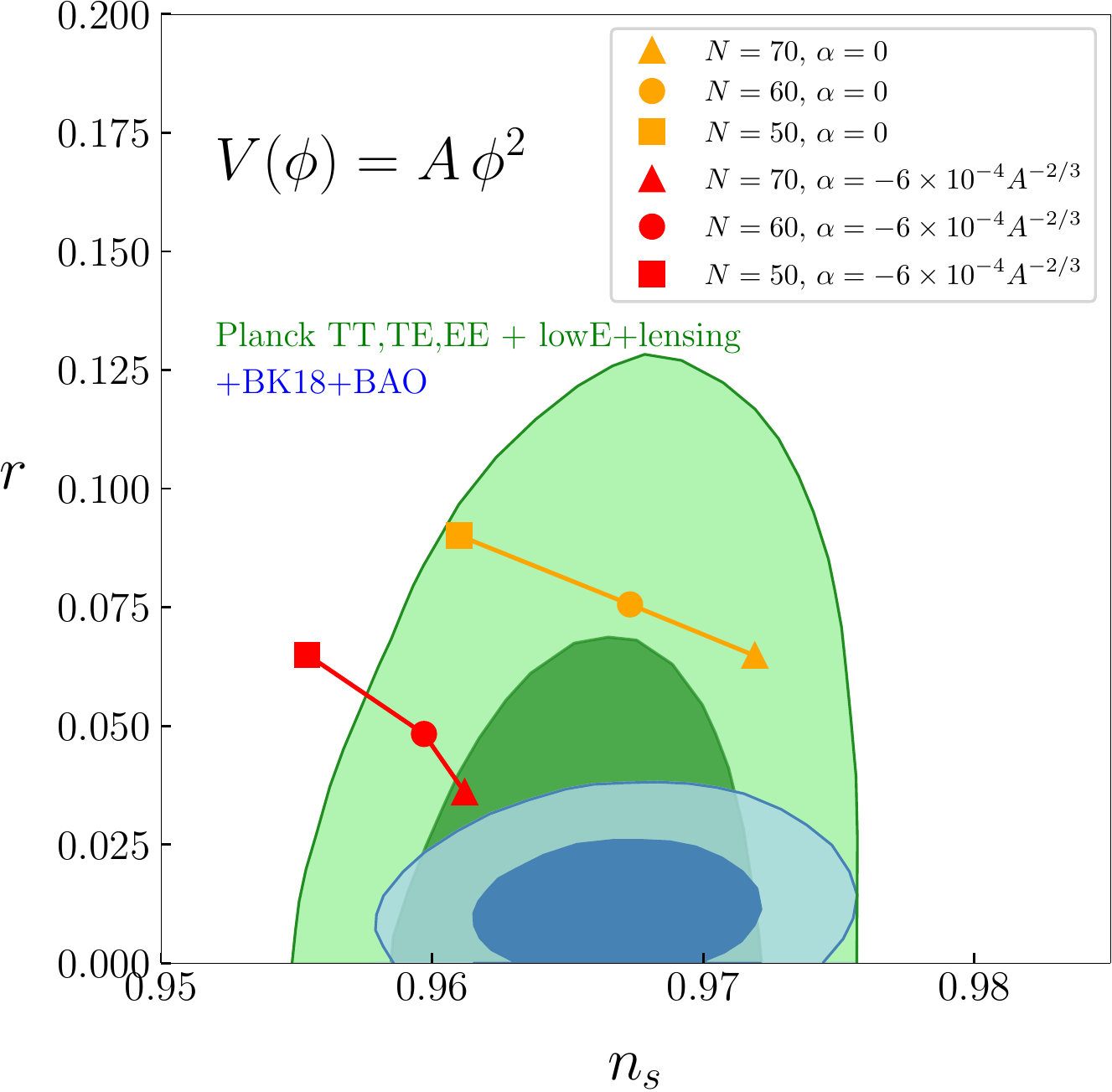}\hspace{0.5cm}
 \includegraphics[scale=0.45]{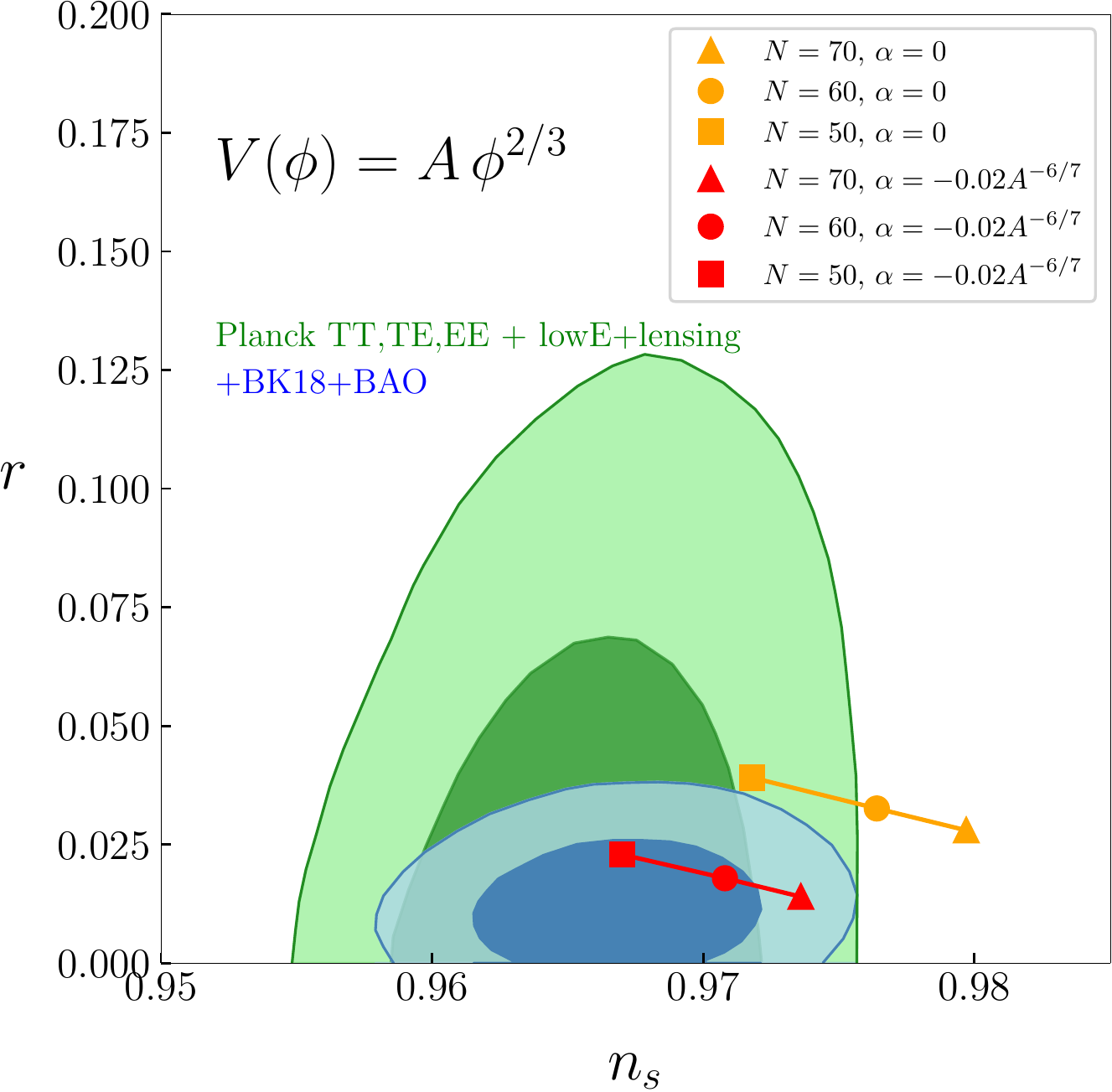}
 \caption{ Tensor-to-scalar ratio vs spectral index for the non-canonical EMSG model with the power law potential $V=A \phi^{n}$ for  $n=2$ (right) $n=2/3$ (left), compared to the data of Ref.~\cite{ade2021improved}.  \label{Fig3}}
\end{figure}

Now let's consider a non-canonical Lagrangian density for a scalar field as~\cite{Li:2012vta}
\begin{equation}\label{Lmnon}
\mathcal{L}_{\rm m}=X^{\delta}-V(\phi)
\end{equation}
where $\delta$ is a positive constant, for which $\delta=1$ we recover the canonical scalar field Lagrangian. The corresponding Lorentz scalar reads
\begin{equation}
\mathbb{T}^2=4\Big[(\delta (\delta-1)+1) X^{2 \delta}+(\delta-2)V X^{\delta}+V^2\Big].
\end{equation}
Here, we apply the slow-roll approximation to evaluate the most important inflationary parameters for the EMSG model with $\beta=1$. In this type of K-essence model, with $P(X,\phi)=X^{\delta}-V -\alpha \mathbb{T}^{2}/2$, we can derive equations for the scale factor $a(t)$ and the scalar field $\phi$, in the slow-roll limit, as follows:\footnote{During the derivation of field equations, it is crucial to recognize that the term $\frac{\partial^2 \mathcal{L}_{\rm m}}{\partial g^{\mu\nu} \partial g^{\sigma\epsilon}}$ exhibits non-zero behavior in the context of non-canonical scalar fields, in contrast to canonical scalar fields. It is important to note that the decision to treat this term as zero remains a matter of choice. For further reading on this topic, refer to Ref.~\cite{us}.}
\begin{equation}\label{Hnon}
3 H^2\simeq V(1+2 \alpha V),
\end{equation}
\begin{equation}\label{vpnon}
V' \dot{\phi}(1+4 \alpha V)\simeq -6 \delta H X^{\delta} \Big(1+2 \alpha (2-\delta) V\Big).
\end{equation}
Taking advantage of Eq.~\eqref{Hnon}, we can calculate the Hubble slow-roll parameter as
\begin{equation}\label{epsilonv}
\varepsilon_{H} \simeq -\frac{1}{2 }\Big(\frac{V'}{V }\Big)\Big(\frac{\dot{\phi}}{H}\Big)\Big(\frac{1+4\alpha V }{1+2\alpha V }\Big).
\end{equation}
From Eq.~\eqref{vpnon}, we can write the ratio $\dot{\phi}/H$ as 
\begin{equation}\label{phidoth1}
\frac{\dot{\phi}}{H} = -K(\delta)\Big[\frac{V'(1+4 \alpha V)}{V^{\delta}\Big(1+2 \alpha V\Big)^{\delta} \Big(1+2 \alpha (2-\delta) V\Big)}\Big]^{\frac{1}{2 \delta-1}},
\end{equation}
where $K(\delta)=\Big(\frac{6^{\delta-1}}{\delta}\Big)^{\frac{1}{2 \delta-1}}$. Now, it is straightforward to verify that
\begin{equation}\label{epsilonv}
\varepsilon_{H} \simeq \frac{K(\delta)}{2 }\left[\frac{\Big(V'(1+4 \alpha V)\Big)^{2 \delta}}{\Big(V(1+2 \alpha V)\Big)^{3\delta -1} }\right]^{\frac{1}{2 \delta-1}} \Big(1+2 \alpha (2-\delta) V\Big)^{\frac{1}{1-2\delta}}.
\end{equation}
Similarly, the sound speed $c_s$ takes the following form
\begin{equation}\label{m1}
c_s^2=\frac{1}{2 \delta-1}+\frac{8 \alpha \delta (1-\delta+\delta^2)}{6^{\delta}(2 \delta-1)^2}\,\frac{\Big(V(1+2 \alpha V)\Big)^{\delta}}{1+4 \alpha V(1- \frac{1}{2} \delta)  \alpha} \Big(\frac{\dot{\phi}}{H}\Big)^{2\delta}
\end{equation}
Under the assumption of small $\alpha V$, we obtain the relation between the potential and the e-folding number $N$ as
\begin{eqnarray}\label{Vexpand2}
V\simeq ( \kappa K N )^{\frac{n}{\kappa}} n^{\frac{n}{\kappa(2 \delta-1)}}A^{\frac{2 \delta}{\kappa (2 \delta-1)}}\Big[1+f_3 \alpha A^{\frac{4 \delta}{\kappa (2 \delta-1)}}
+ f_4 \alpha^2 A^{\frac{6 \delta}{\kappa (2 \delta-1)}}+ \mathcal{O}(\alpha^{3}A^{\frac{8 \delta}{\kappa (2 \delta-1)}})\Big],
\end{eqnarray}
where the new functions $f_3(N)$ and $f_4(N)$ are defined as
\begin{equation}
\begin{split}
&f_3(N)=2 (\kappa K N)^{\frac{2 n}{\kappa}}\Big(\frac{ \delta (\delta-3)}{2 \delta+n (5 \delta-3)}\Big)n^{1+\frac{2 n}{\kappa(2 \delta-1)}},\\&
f_4(N)=8 (\kappa K N)^{\frac{3 n}{\kappa}}\frac{ \delta (11+(\delta-6)\delta)}{3(2 \delta+n (7 \delta-4))}n^{1+\frac{3 n}{\kappa(2 \delta-1)}}
\end{split}
\end{equation}
with $\kappa(\delta,n)=\frac{n(\delta-1)+2\delta}{2 \delta-1}$. Then, using Eq.~\eqref{epsilonv} and Eq.~\eqref{Vexpand2}, we reach the final relation for the spectral index and the tensor to scalar ratio in Eqs.~\eqref{nsanalatic}  and~\eqref{rformula}; 
\begin{eqnarray}\label{rhons}
 &n_{s}&-1 \simeq  -\frac{1}{N}\Big[\mathcal{I}+ (K N)^{\frac{2 n}{\kappa}}\mathcal{N} \alpha^2 A^{\frac{4\delta}{\kappa (2\delta-1)}}+\mathcal{O}(\alpha^{3}A^{\frac{8 \delta}{\kappa (2 \delta-1)}}) \Big],\\
\label{rhor} r&\simeq & 16 \varepsilon_{H} c_{s} \simeq \frac{8 n }{N \kappa}\Big[c_{s}+f_5 \alpha A^{\frac{2 \delta}{\kappa(2 \delta-1)}}
 + f_6\alpha^2 A^{\frac{4 \delta}{\kappa(2 \delta-1)}} +\mathcal{O}(\alpha^{3}A^{\frac{8 \delta}{\kappa (2 \delta-1)}})\Big],
\end{eqnarray}
where
\begin{equation}
\mathcal{I}=\frac{n(3 \delta-2)+2\delta}{n(\delta-1)+2 \delta}
\end{equation}
where $\mathcal{N}$ is a complicated polynomial function of $\delta$ and $n$, which we cannot present here. However, for the chaotic potential ($n=2$), it simplifies to:
\begin{equation}
\mathcal{N}=2^{\frac{2(4 \delta-1)}{2 \delta-1}}\,\frac{6+\delta(\delta-15)}{6 \delta-3}.
\end{equation}
On the other hand, the functions $f_5(N)$ and $f_6(N)$ are found to be
\begin{equation}
\begin{split}
&f_5(N)=\frac{1}{8} (\kappa K N)^{\frac{n}{\kappa}} n^{\frac{n}{\kappa(2\delta-1)}},\\&
f_6(N)=\frac{1}{4} (\kappa K N)^{\frac{2 n}{\kappa}} n^{\frac{2n}{\kappa(2\delta-1)}}\,\frac{n(3+\delta(\delta-8))-2 \delta}{2 \delta +n (5 \delta-3)}.
\end{split}
\end{equation}
Under the assumption of slow-roll, using~\eqref{m1}, the sound speed squared is found to be
\begin{eqnarray}\label{rhocs}
c_{s}^2=
\frac{1}{2 \delta-1}+ f_7 \alpha A^{\frac{2 \delta}{\kappa(2 \delta-1)}}+
\mathcal{O}( \alpha^2 A^{\frac{4 \delta}{\kappa(2 \delta-1)}}).
\end{eqnarray}
We find that
\begin{equation}
\begin{split}
f_7(N)=\frac{8\delta}{6^{\delta}} K^{\frac{(n+2)\delta}{\kappa}} (n N)^{\frac{(2+n)\delta}{\kappa (2 \delta-1)}} (2 \delta-1)^{\frac{2n-2\delta(2+n)}{\kappa(2 \delta-1)}} (1+ \delta (\delta-1)) \kappa^{\frac{(n-2) \delta}{\kappa (2 \delta-1)}}
\end{split}
\end{equation}
as $\alpha \to 0$, which is in consistent with one given in~\cite{Li:2012vta}. And, from~\eqref{lamda}, we find that
\begin{equation}
\frac{\lambda}{\Sigma} \simeq \frac{1}{3} (\delta-1) + \frac{(4 \delta-1)}{6(2 \delta-1)} \Big[1-2 \delta +\frac{1}{c_{s}^2}\Big].
\end{equation}
Finally, the non-Gaussianity parameter $f^{\rm{equi}}_{\rm{NL}}$~\eqref{fNl1} at leading order is obtained to be
\begin{equation}
f^{\rm{equi}}_{\rm{NL}}\simeq \frac{85}{324} \Big(1-\frac{1}{c_{s}^2}\Big)-\frac{40}{324} \Big(\frac{1}{3} (\delta-1) + \frac{(4 \delta-1)}{6(2 \delta-1)} \Big[1-2 \delta +\frac{1}{c_{s}^2}\Big]\Big).
\end{equation}
Note that as one takes $c_{s}^{-2}=2 \delta-1$, this  matches with ~\cite{unnikrishnan2013resurrecting} for $\alpha \to 0$. For instance, in EMPG model, by selecting a chaotic potential $V=A \phi^2$ and $\mathcal{L}=X^2-V$, the non-Gaussianity parameter reads
\begin{eqnarray}
\nonumber f_{\rm{NL}}^{\rm{equi}} &\simeq & -\frac{275}{486}+\frac{1810}{243} \Big(\frac{2}{9} A^{2}\Big)^{\frac{1}{3}} \alpha  +\mathcal{O}\Big((\alpha^{2} A^{4/3})\Big) \simeq -\frac{275}{486},
\end{eqnarray}
This result is in excellent agreement with the Planck result of $f^{\rm{equi}}_{\rm{NL}}=-26\pm 47$ ($68 \%$ C.L.)~\cite{Planck:2018jri}, as confirmed by the numerical data presented in Tabels~\ref{table3} and~\ref{table4}. Our view is that the results reported in the tables emphasize the validity of our analytical solutions~\eqref{rhocs},~\eqref{rhons}, and~\eqref{rhor} in the slow-roll limit for inflationary parameters. The most remarkable finding in our scenario is that the inclusion of the $\mathbb{T}^2$ term in the non-canonical Lagrangian leads to a shift of the data on $r$ and $n_{s}$ towards the BICEP/Keck (BK15 and BK18) plus BAO bound. In other words, as shown in Fig.~\ref{Fig3}, the values of the inflationary parameters $n_{s}$ and $r$ reported in Ref.~\cite{Li:2012vta} are improved in non-canonical EMSG inflation, making them compatible with current observational constraints~\cite{ade2021improved}.

	\begin{table*}
	\centering
	\begin{tabular}{p{1.9cm} ||p{1.2cm}p{1.2cm}p{1.2cm}p{1.2cm}||p{1.2cm}p{1.2cm} p{1.1cm}p{1.2cm}}
 \hline 
 $N=70$ &  &  \textbf{Numerics}  &                              &   &       &      \textbf{Analytics}                  &                           &                  \\
	\hline \hline
	$|\alpha| A^{2/3}$                               &$n_{\text{s}}$       &$ r$                     &$ c_{s}^{2}$    &$|f^{\rm equil}_{\rm{NL}}|$          &$n_{\text{s}}$       &$r$                          &$ c_{s}^{2}$            &$|f^{\rm equil}_{\rm{NL}}|$\\
	\hline
   $8.5 \times 10^{-4}$               &$0.9253$               &$0.0170$             &$0.3331$           &$0.5663$        &$0.9271$               &$0.0143$                  &$0.3332$                  &$0.5667$\\    
	\hline
   $7.9 \times 10^{-4}$                &$0.9391$              &$0.0212$              &$0.3330$          &$0.5665$          &$0.9405$              &$0.0193$                 &$0.3331$                  &$0.5666$\\
	\hline
	$7.1 \times 10^{-4}$               &$0.9513$              &$0.0271$              &$03330$            &$0.5666$         &$0.9520$              &$0.0262$                  &$0.3330$             &$0.5663$\\ 
	 \hline
	$6.6\times 10^{-4}$                &$0.9568$              &$0.0316$              &$0.3329$           &$0.5667$         &$0.9569$              &$0.0310$                  &$0.3330$                   &$0.5661$\\                                                                                                                   
    \hline   
	$6.0 \times 10^{-4}$               &$0.9612$               &$0.0360$             &$0.3329$            &$0.5668$        &$0.9610$               &$0.0356 $                 &$0.3329$              &$0.5659$\\                                                                                            
	\hline \hline
 $N=60$                      &                              \cellcolor{gray} &                            \cellcolor{gray} &                        \cellcolor{gray} &                      \cellcolor{gray} &                          \cellcolor{gray} &                             \cellcolor{gray} &                     \cellcolor{gray} &  \cellcolor{gray}                       \\
	\hline \hline 
     $8.5\times 10^{-4}$                 &$0.9412$            &$0.0311$               &$0.3329$            &$0.5668$      &$0.9418$                  &$0.0297$                 &$0.3330$               &$0.5667$\\ 
	 \hline
  $7.9 \times 10^{-4}$                  &$0.9478$              &$0.0353$              &$0.3329$             &$0.5669$       &$0.9481$                &$0.0344$                &$0.3329$               &$0.5668$\\    
    \hline
	$7.1 \times 10^{-4}$                &$0.9542$             &$0.0411$               &$0.3329$               &$0.5670$     &$0.9539$                 &$0.0405$                 &$0.3329$               &$0.5669$\\ 
	 \hline
	$6.6 \times 10^{-4}$                &$0.9570$             &$0.0441$               &$0.3329$               &$0.5670$     & $0.9556$                 &$0.0438$                &$0.3329$                   &$0.5670$\\                                                                                                       
\hline
	$6.0 \times 10^{-4}$                &$0.9597$             &$0.0483$               &$0.3328$                  &$0.5670$    &$0.9591$                  &$0.0481$                 &$0.3328$                    &$0.5670$\\                                                                                                         
	\hline 
		\end{tabular}
		\caption{A comparison between analytical and numerical results for inflationary parametersfor inflationary parameters in the non-canonical EMSG model with $\delta=2$ by considering a chaotic potential, $V=A \phi^{2}$. \label{table3} }
\end{table*}

	\begin{table*}
	\centering
	\begin{tabular}{p{1.9cm} ||p{1.2cm}p{1.2cm}p{1.2cm}p{1.2cm}||p{1.2cm}p{1.2cm} p{1.1cm}p{1.2cm}}
 \hline 
 $N=70$ &  &  \textbf{Numerics}  &                              &   &       &      \textbf{Analytics}                  &                           &                  \\
	\hline \hline
	$|\alpha| A^{6/7}$         &$n_{\text{s}}$        &$ r$                     &$ c_{s}^{2}$    &$|f^{\rm equil}_{\rm{NL}}|$          &$n_{\text{s}}$       &$r$                          &$ c_{s}^{2}$            &$|f^{\rm equil}_{\rm{NL}}|$\\
	\hline
   $2.5 \times 10^{-2}$    &$0.9629$                &$0.0086$            &$0.3332$            &$0.5661$       &$0.9629$               &$0.0081$                 &$0.3332$                &$0.5660$\\    
	\hline
   $2.0 \times 10^{-2}$     &$0.9736$               &$0.0140$            &$0.3331$            &$0.5662$         &$0.9731$              &$0.0138$                &$0.3331$                  &$0.5662$\\
	\hline
	$1.5 \times 10^{-2}$     &$0.9776$              &$0.0185$            &$0.3331$             &$0.5663$      &$0.9767$                 &$0.0184$                &$0.3331$              &$0.5663$\\ 
	 \hline
	$1.0\times 10^{-2}$      &$0.9790$              &$0.0224$              &$0.3331$           &$0.5662$       &$0.9781$                &$0.0223$                  &$0.3331$              &$0.5662$\\                                                                                                                        
\hline
    $0$	                              &$0.9797$              &$0.0280$              &$0.3333$          &$0.5658$         &$0.9787$               &$0.0279$                  &$0.3333$                    &$0.5658$\\
	\hline \hline
    $N=60$                      &                              \cellcolor{gray} &                            \cellcolor{gray} &                        \cellcolor{gray} &                      \cellcolor{gray} &                          \cellcolor{gray} &                             \cellcolor{gray} &                     \cellcolor{gray} &  \cellcolor{gray}                       \\
	\hline \hline 
    $2.5\times 10^{-2}$       &$0.9625$            &$0.0118$              &$0.3332$            &$0.5662$          &$0.9621$                          &$0.0114$                 &$0.3332$                &$0.5661$\\
	\hline
   $2.0\times 10^{-2}$        &$0.9708$             &$0.0179$              &$0.3331$           &$0.5663$           &$0.9700$                         &$0.0177$                 &$0.3331$                &$0.5663$\\ 
	 \hline
  $1.5 \times 10^{-2}$         &$0.9743$            &$0.0226$               &$0.3331$           &$0.5663$          &$0.9733$                          &$0.0224$                 &$0.3331$               &$0.5663$\\    
    \hline
	$1.0 \times 10^{-2}$       &$0.9756$             &$0.0267$              &$0.3331$            &$0.5663$          &$0.9746$                          &$0.0265$                  &$0.3331$                   &$0.5663$\\ 
	 \hline
	$0$                                  &$0.9764$             &$0.0326$              &$0.3333$            &$0.5658$          &$0.9752$                           &$0.0325$                &$0.3333$              &$0.5658$\\                                                                                                       
 \hline 
		\end{tabular}
		\caption{A comparison between analytical and numerical results for inflationary parameters for inflationary parameters in the non-canonical EMSG model with $\delta=2$ by considering the potential, $V=A \phi^{\frac{2}{3}}$. \label{table4}}
\end{table*}

As a final remark, let us discuss the prediction of the inflationary parameters in the scale-independent EMSG ($\beta=1/2$) under the slow-roll scheme. The dynamical background equations and the sound speed squared are obtained as
\begin{eqnarray}
3 H^2 &\simeq & (1+\alpha) V\quad,\\ 
V' \dot{\phi}
 &\simeq & -\frac{6\delta H X^{\delta}}{1+\alpha} \Big(1-\frac{\alpha}{2}(\delta-2)\Big)\quad,\\ \label{csbeta12}
c_{s}^2&\simeq & \frac{1}{2 \delta-1}+\frac{3\alpha \delta^3 (\alpha+1)^{\delta}}{6^{\delta}(2\delta-1)^2 (2-\alpha(\delta-2))} V^{\delta-1} \Big(\frac{\dot{\phi}}{H}\Big)^{2 \delta}\quad,\\
\label{epsilon12} \varepsilon_{H}&=&-\frac{1}{2} \Big(\frac{V'}{V}\Big) \frac{\dot{\phi}}{H}\quad, \quad \eta_{H}= \frac{\dot{\varepsilon_{H}}}{H \varepsilon_{H}}= \frac{\varepsilon_{H}'}{ \varepsilon_{H}} \frac{ \dot{\phi}}{ H}\quad.
\end{eqnarray}
Note that if we choose $\delta=2$, the last term in the sound speed relation vanishes. Using the first two equations above, we can derive the ratio $\dot{\phi}/H$ as follows:
\begin{equation}
\frac{\dot{\phi}}{H}=-K(\delta) \Big[\frac{(1+\alpha)^{1-\delta}}{1-\frac{\alpha}{2}(\delta-2)} \Big(\frac{V'}{V^{\delta}}\Big)\Big]^{\frac{1}{2 \delta-1}}.
\end{equation} 
Hence, in the power law potential case $V=A \phi^{n}$, it is straightforward to confirm that 
\begin{equation}
N=\frac{n}{\kappa(\delta,n)}\frac{1}{\chi(\delta,n) } V^{\kappa(\delta,n)/n}; \hspace{0.5cm} \chi(\delta,n)=K(\delta) \Big[\frac{(1+\alpha)^{1-\delta}}{1-\frac{\alpha}{2}(\delta-2)}\Big]^{\frac{1}{2 \delta-1}} \Big(n A^{\frac{1}{n}}\Big)^{\frac{2 \delta}{2 \delta-1}}.
\end{equation}
We can establish a relationship between the sound speed, slow-roll parameters, and $N$ by combining Eqs.~\eqref{csbeta12},~\eqref{epsilon12}, and the equation above;
\begin{eqnarray}
\label{csbeta12rho}c_{s}^2 &=& \frac{1}{2 \delta-1}+\frac{n\alpha(1+\alpha)(-1)^{2\delta} \delta^2}{\kappa (\alpha(\delta-1)-2)^2 (2 \delta-1)^2}\frac{1}{N}\quad,\\
\epsilon_{H}&=&\frac{\delta}{2 \kappa} \frac{1}{N}\quad, \quad \eta_{H}= \frac{1}{N}.
\end{eqnarray}
Accordingly, the inflationary parameters, $n_{s}$, $r$, and $f_{\rm NL}^{\rm equil}$ read
\begin{eqnarray}
n_{s}-1&=&-\mathcal{I}(\delta,n) \frac{1}{N}\quad\,\\
r &\approx & 16 \varepsilon_{H} c_{s} = \frac{8 n}{\kappa(2\delta-1) N}\Big[1+\frac{(-1)^{2 \delta}\alpha (\alpha+1) \delta^2}{\kappa (\alpha(\delta-2)-2)^2 (1-2 \delta) N}\Big]\quad\,\\
f_{\rm NL}^{\rm equil}&=& \frac{1}{972 (2 \delta-1) }\Big[(275-590 \delta)\Big(\frac{1}{c_{s}^2}-1\Big)+80 \delta (\delta-1)\Big].
\end{eqnarray}
As $\alpha$ approaches zero, these findings align with the values presented in~\cite{unnikrishnan2013resurrecting, Li:2012vta}. Based on Eq.~\eqref{csbeta12rho}, the negative value of $\alpha$ results in smaller $c_s^2$ values (less than 1/3). As a result, $f_{\rm NL}^{\rm equi}$ is lower compared to the $\alpha=0$ scenario~\cite{unnikrishnan2013resurrecting}. While other inflationary parameters, such as $n_s$ and $r$, match well with those reported in~\cite{Li:2012vta}.

\section{Conclusions}\label{consec}
In this paper, we have studied a single scalar field inflation within the framework of a particular form of energy momentum squared gravity (EMSG ) with an extra piece, $f(\mathbb{T}^2)=-\alpha \mathbb{T}^{2\beta}$ added to Einstein-Hilbert action.
As stated  in the introduction, the EMSG theory has been investigated in different contexts. For example, in~\cite{roshan2016energy}, a cosmological bouncing solution was found for the specific cases of $\beta=1$ and $\alpha>0$. Another instance is the scale-independent EMSG ($\beta=1/2$) with $\alpha=2$, which can reproduce the original steady state universe in the presence of dust, as demonstrated in~\cite{akarsu4}. It is worth noting that our analysis reveals that both models are subject to instabilities due to the positive sign of the coupling parameter $\alpha$.

We have shown that, in case $\mathbb{T}^2$ is constituted by a scalar field, the EMPG is equivalent to a specific K-essence model. Subsequently, we have examined the tensor and scalar perturbations of the metric around the FRW background in the presence of an inflaton scalar field $\phi$. We have shown that to circumvent the ghost and gradient instabilities, the sound speed and the slow roll parameter $\varepsilon_H$ must be positive. Consequently, we have discussed the constraints on the free parameter $\alpha$. Our findings indicate that, for any value of $\beta$, the $\alpha$ parameter should be negative ($\alpha<0$). And, in the particular case $\beta=1/2$, i.e., the scale-independent EMSG, we must have $-2<\alpha<0$. Then, we have examined the slow roll inflation in EMPG using two different matter Lagrangians; the canonical scalar field case ($\mathcal{L}_{\rm m}=X-V(\phi)$), and the non-canonical scalar field case ($\mathcal{L}_{\rm m}=X^{\delta}-V(\phi)$) with some specific forms of the field potential. We have shown the presence of EMPG modifications allows us to bring the parameters $r$ and $n_{s}$ in consistency with the recent BICEP/Keck constraints on these parameters in both the canonical and non-canonical scalar field cases. Furthermore, we have found that in the case of canonical Lagrangian, the $f^{\rm equil}_{\rm{NL}}$ takes small values similar to those of standard chaotic inflation. In contrast, for the non-canonical Lagrangian case, the $f^{\rm equil}_{\rm{NL}}$ turns out to be larger.

As a follow-up to the present investigations, it would be interesting to examine the possibility of large fluctuations in the framework of EMSG and associated issues related to primary and secondary GWS \cite{Li:2023qua,Bodas:2022urf,Chen:2022dah,Cicoli:2022sih,Lin:2021vwc,Rezazadeh:2021clf,Ahmed:2021ucx,Wu:2021zta,Choudhury:2013woa,Inomata:2019ivs,Wang:2019kaf,Alabidi:2012ex}. Unfortunately, the scope of PBH formation in this case is limited by quantum loop effects \cite{Kristiano:2022maq,Riotto:2023hoz,Choudhury:2023vuj,Choudhury:2023jlt,Kristiano:2023scm,Riotto:2023gpm,Choudhury:2023rks,Choudhury:2023hvf,Firouzjahi:2023ahg,Kawaguchi:2023mgk,Cheng:2023ikq,Tasinato:2023ukp,Franciolini:2023lgy,Motohashi:2023syh,Firouzjahi:2023aum}.

While our primary focus in this paper has been on inflation, it is worth mentioning in closing that our specific K-essence model might also hold significance for the study in the context of the late universe, in relevance with dark energy and/or cold dark matter. Such behaviors in scalar fields have been previously documented in \cite{de}. Conducting a dynamical system analysis could offer valuable insights into diagnosing the model and identifying the existence of phases dominated by dark energy and dark matter in the cosmic history of this model, as explored in \cite{kashfi}. We defer these for future investigations.

\acknowledgments
The authors thank Hassan Firouzjahi, Shahab Shahidi, Zahra Haghani, Sayantan  Choudhury, Mohamad Ali Gorji, Alireza Talebian, and Phongpichit  Channuie for the useful comments and discussions. SAH and FF acknowledge the partial support from the “Saramadan” federation of Iran. The research of MR is supported by the Ferdowsi University of Mashhad. \"{O}.A. acknowledges the support by the Turkish Academy of Sciences in scheme of the Outstanding Young Scientist Award  (T\"{U}BA-GEB\.{I}P). MS is supported by Science and Engineering Research Board (SERB), DST, Government of India under the Grant Agreement number CRG/2022/004120 (Core Research Grant). MS is also partially supported by the Ministry of Education and Science of the Republic of Kazakhstan, Grant No. AP14870191 and CAS President's International Fellowship Initiative(PIFI). 

\bibliographystyle{JHEP}
\bibliography{Tinflation1}

\end{document}